\documentclass[12pt,
preprintnumbers,amsmath,amssymb,nofootinbib]{revtex4}

\usepackage{amsmath,amsfonts}
\usepackage{graphicx}
\usepackage{psfrag}

\newcommand{\BC}{{\mathbb C}}
\newcommand{\BZ}{{\mathbb Z}}
\newcommand{\CP}{{\mathbb CP}}
\newcommand{\CO}{{\cal O}}

\newcommand{\ket}[1]{{|#1 \rangle}}
\newcommand{\BR}{{\mathbb R}}

\newcommand{\diag}{\hbox{diag}}
\newcommand{\bea}{\begin{eqnarray}}
\newcommand{\eea}{\end{eqnarray}}
\renewcommand{\bar}{\overline}

\newcommand{\tr}{\hbox{Tr}}
\newcommand{\vac}{|0\rangle}

\begin{document}
\title{ Three-dimensional ${\cal N}=6$ SCFT's  and their membrane dynamics}
\author{David Berenstein}
\email{dberens@physics.ucsb.edu}
\author{Diego Trancanelli}
\email{dtrancan@physics.ucsb.edu}

 \affiliation{Department of
Physics, University of California
at Santa Barbara, CA 93106}

\begin{abstract}
\noindent
We analyze several aspects of the recent construction of three-dimensional conformal gauge theories by Aharony {\it et al.} in various regimes. We pay special attention to understanding how the M-theory geometry and interpretation can be extracted from the analysis of the field theory. We revisit the calculations of the moduli space of vacua and the complete characterization of chiral ring operators by analyzing the field theory compactified on a 2-sphere. We show that many of the states dual to these operators can be interpreted as D-brane states in the weak coupling limit. Also, various features of the dual $AdS$ geometry can be obtained by performing a strong coupling expansion around moduli space configurations, even though one is not taking the planar expansion. In particular, we show that at strong coupling the corresponding  weak coupling D-brane states of the chiral ring localize on particular submanifolds of the dual geometry that match the M-theory interpretation. We also study the massive spectrum of fields in the moduli space. We use this to investigate the dispersion relation of giant magnons from the field theory point of view. Our analysis predicts the exact functional form of the dispersion relation as a function of the world-sheet momentum, independently of integrability assumptions, but not the exact form with respect to the 't Hooft coupling. We also get the dispersion relation of bound states of giant magnons from first principles, providing evidence for the full integrability of the corresponding spin chain model at strong 't Hooft coupling.

\end{abstract}

\maketitle


\section{Introduction}

Finding a Lagrangian formulation for the conformal field theory dual to M-theory on $AdS_4 \times S^7$ has been a long-standing problem in the AdS/CFT correspondence. Major progress in this direction has been recently triggered by the construction, based on three-algebras, by Bagger, Lambert, and Gustavsson (BLG) \cite{Bagger:2006sk,Gustavsson:2007vu} of the world-volume action for two M2-branes. This progress has later prompted Aharony, Bergman, Jafferis, and Maldacena (ABJM) \cite{ABJM} to propose  that a Chern-Simons action coupled to an appropriately chosen matter sector\footnote{Superconformal extensions of Chern-Simons theory with various amounts of supersymmetry have been also considered recently in \cite{Schwarz:2004yj,Gaiotto:2007qi,Gaiotto:2008sd,Hosomichi:2008jd}.} generalizes the BLG construction to an arbitrary number of coincident M2-branes, without incurring in the major shortcomings of the BLG model. The ABJM model is manifestly unitary and generalizes to any number of branes without difficulty.

This achievement comes at the cost of sacrificing manifest R-symmetry, so that the ABJM theory has manifestly only an $SU(4)_R$ global symmetry and it is ${\cal N}=6$ supersymmetric.\footnote{The supersymmetries of this action have been checked carefully in \cite{Hosomichi:2008jb,Bandres:2008ry}, while its relation to the construction in terms of three-algebras has been described in \cite{Bagger:2008se,Kim:2008gn}.}

More precisely, the ABJM theory consists of a $U(N)\times U(N)$ Chern-Simons gauge field,\footnote{A classification of ABJM-like theories with ${\cal N}=5,6$ supersymmetries can be found in \cite{Hosomichi:2008jb,Schnabl:2008wj}, while the analysis of the $U(N)\times U(M)$ case and of orthogonal and symplectic groups was carried over in \cite{Hosomichi:2008jb,Aharony:2008gk}.} with levels $k$ and $-k$, coupled to two pairs of chiral superfields transforming in bifundamental representations of the gauge group. We call them $\phi^A=(A_a,\bar B_{\dot a})$ with $A=1,\ldots,4$ and $a,\dot a=1,2$ (see the appendix for more details about conventions), so that $A_a$ transforms in the $(N,\bar N)$ while $B_{\dot a}$ in the $(\bar N,N)$. In the dual geometry, the rank of the gauge groups $N$ is interpreted as the units of flux of the four-form threading the sphere, while the Chern-Simons level $k$ specifies the orbifold singularity at which the M2-branes sit: the transverse space is in fact given by $\mathbb{C}^4/\mathbb{Z}_k$, or, after the near horizon limit, by $S^7/\mathbb{Z}_k$.

The ABJM theory, unlike \cite{Bagger:2006sk,Gustavsson:2007vu},  enjoys a large $N$ limit and one can also define a 't Hooft coupling constant, given by $\lambda\equiv N/k$.  For $k \gg N$ the theory is weakly coupled, while for $N\gg 1$ and $k\ll N$ it has a gravity dual which is either M-theory on $AdS_4\times S^7/\mathbb{Z}_k$ (if $k\ll N^{1/5}$) or type IIA strings on $AdS_4\times \mathbb{C}P^3$ (if $N^{1/5}\ll k\ll N$) \cite{ABJM}. Taking a large $k$ shrinks in fact the radius of the M-theory circle fibered over the $\mathbb{C}P^3$ base, thus reducing the dimensions of the space from eleven to ten.

A notable feature of the string theory limit of the ABJM theory, whose sigma model was studied in \cite{Arutyunov:2008if,Stefanski:2008ik} and, using a covariant formulation, in \cite{D'Auria:2008cw},  is that it appears to be integrable \cite{Minahan:2008hf,Gromov:2008bz} (see also \cite{Ahn:2008aa,Bak:2008cp}). A flurry of activity has followed this observation. In particular, the Penrose limit of the $AdS_4\times \mathbb{C}P^3$ background was considered in  \cite{Nishioka:2008gz,Gaiotto:2008cg,Grignani:2008is}
and the ${\cal N}=6$ Chern-Simons spin chain was studied in \cite{Gaiotto:2008cg,Grignani:2008is,Bak:2008cp}.\footnote{Semiclassical strings and one-loop corrections to their energies were discussed in  \cite{Chen:2008qq} and \cite{McLoughlin:2008ms,Gromov:2008fy}.}

In this paper we address several aspects of the ABJM proposal. These range from providing arguments that this is really a theory of membranes, rather than of D-branes, to analyzing its moduli space of vacua and its chiral ring both at weak and strong coupling, and to proving exactly  the functional form for the dispersion relation of the giant magnon in the ${\cal N}=6$ spin chain (in some sense this is a resummation to all orders in perturbation theory).
We give more details about the organization of our results next, in the outline section. 

The main tool we have used is a semiclassical approximation around classical configurations. These are given by eigenvalues of a matrix quantum mechanics obtained by dimensionally reducing the theory on $S^2\times \mathbb{R}$. Many of these configurations should be considered as non-perturbative saddles, so that we are exploring the ABJM model beyond 
perturbative field theory around a trivial vacuum.

The off-diagonal modes  connecting pairs of eigenvalues are interpreted as fluctuations around the classical background. Their mass plays a crucial role in many of our computations. Differently from ${\cal N}=4$ SYM, which has a quartic scalar potential and consequently a mass of off-diagonal modes which scales like the distance between the eigenvalues, ${\cal N}=6$ Chern-Simons has a sextic potential and masses that scale like the distance squared. We will see how this fact has profound consequences on the physics of the theory.


\subsubsection*{Outline}

We summarize here the content of each one of the sections that compose this paper.

In Section \ref{sec:M2vsD2} we present evidence in support of the claim that the ABJM theory describes M2- rather than D2-branes.\footnote{Some work addressing the relation between M2- and D2-branes in this context can be found in \cite{Li:2008ya,Pang:2008hw, Verlinde:2008di}.} We comment in particular on the differences between standard orbifold constructions for D-branes \cite{DM} and the action of the orbifold in the ABJM theory, which has the effect to change the level of the Chern-Simons coupling.\footnote{Orbifolds for the BLG and ABJM models are also considered in \cite{Terashima:2008ba,Kim:2008gn}.} We study the off-diagonal modes connecting parallel M2-branes and show that they can be interpreted as extended objects whose tension goes like $\ell_P^{-3}$, so they behave like a two-dimensional object rather than like a string. We call these {\it membrane bits}.  We also introduce an explicit definition of Wilson loop operators in terms of the off-diagonal modes stretching between membranes and emphasize how they differ from the D-brane case. 

In Section \ref{sec:dimred} we start the analysis at the classical level of the dimensional reduction of the ABJM theory on $S^2\times \mathbb{R}$. Requiring spherically symmetric configurations, we find that the $U(N)\times U(N)$ gauge field is generically  broken down to $U(1)^N\times U(1)^N$ by the gauge flux, and that it  induces a flux quantization condition already at the classical level. The bifundamental scalars are charged under these groups and can only acquire spherically symmetric vevs if $U(1)^N\times U(1)^N$ gets further reduced to a diagonal $U(1)^N$, and the fluxes of the two groups are paired by the vevs of the scalars. This type of structure does not depend strongly on the details of the field theory.

In Section \ref{sec:ffBPS} we consider the free field regime of the ABJM theory and discuss BPS operators and their description in terms of droplets of free fermions and Young tableaux. We also study giant and dual giant gravitons, and their excitations. To be concrete, one interesting representative in this family of operators is the maximal giant (the {\it dibaryon}) given by two D4-branes intersecting on a $\mathbb{C} P^1\subset \mathbb{C}P^3$. We discuss the different kinds of strings stretching between these branes and the excitations of the gauge theory operator they correspond to. 

In Section \ref{sec:cr} we study the moduli space of vacua of the ABJM theory. We find that the classical solution to the moduli space problem is in terms of commuting matrices that can be diagonalized simultaneously, so that the dynamics of the system can be analyzed in terms of the eigenvalues of these matrices. We also comment on how choosing two different Chern-Simons levels for the two gauge groups affects the chiral ring states and how one introduces the duals of D0-branes in the field theory setup.

In Section \ref{sec:eg} we consider the proper semiclassical quantization of the modes of the dimensionally reduced theory for the classical chiral ring configurations. This requires to carefully compute the Vandermonde determinant originating from the diagonalization of the scalars. We write the effective Hamiltonian for the system and find that this determinant introduces a repulsive force which makes the eigenvalues localize on a $S^7$ of radius $r_0=\sqrt{2 N}$. We also compute the masses of the off-diagonal degrees of the freedom, finding that, unlike the ${\cal N}=4$ SYM case, where they are proportional to the distance between the eigenvalues that they connect, these are in this instance proportional to the square of the distance from the origin.  

In Section \ref{sec:strongcoup} we look for the geometric objects that are dual at strong coupling to the chiral ring wave functions worked out in Section \ref{sec:ffBPS}. We find, for example, a geometric description of giant gravitons similar to the classification in terms of holomorphic curves found by Mikhailov \cite{Mikhailov:2000ya} for the ${\cal N}=4$ SYM case.
Similarly, we can address the D0-brane systems and find evidence for tachyons in the D0/$\bar{\mbox{D}0}$ system.

In Section \ref{disp-rel-sec} we study the dispersion relation for the giant magnon solution of type IIA string theory on $AdS_4\times \BC P^3$. Using the emergent geometry approach that we have developed in the previous sections, we can easily prove in perturbation theory around the eigenvalue distribution (this one is fixed non-perturbatively) that the exact functional form that was conjectured in \cite{Gaiotto:2008cg,Grignani:2008is} assuming integrability of the ${\cal N}=6$ Chern-Simons spin chain is recovered. Our derivation is independent of the integrability assumption and based solely on the distribution of eigenvalues of the dimensionally reduced theory. This distribution can be computed exactly at weak coupling, and there we find exact agreement with the perturbation theory calculation based on the one loop spin chain dynamics.

We conclude with some comments and a discussion of important open questions in Section~\ref{sec:concl} and collect in the appendix some of the conventions we have used in the paper.


\section{Membranes and not D-branes}
\label{sec:M2vsD2}

Perhaps one of the most surprising aspects of the BLG  \cite{Bagger:2006sk,Gustavsson:2007vu}
and the ABJM \cite{ABJM} theories is that they describe multiple M2-branes as opposed to multiple D2-branes. In this section we present various arguments in support of this statement. We will concentrate on the ABJM model, as this describes an arbitrary number of M2-branes in a natural way, whereas the BLG formulation in terms of three-algebras seems to have a hard time doing this generalization.

The first thing to consider is the case of a single membrane. In its bosonic sector, this theory has a $U(1)\times U(1)$ Chern-Simons gauge field with level $(k,-k)$, and four complex scalar fields, $\phi^A$ ($A=1,\ldots,4$). There is no potential and no Yukawa coupling for the scalars, which are then essentially free. The Chern-Simons action induces a $\BZ_k$-quotient, so that the moduli space is $\mathbb{C}^4/\BZ_k$ \cite{ABJM} (see also \cite{DMPV}). 

It should be noted that in a theory of a single D-brane at a $\BZ_k$ orbifold, we would expect, following the rules of Douglas and Moore for orbifold constructions \cite{DM}, to see a $U(1)^k$ gauge group instead. We would also expect to see strings stretching between D-branes and their images, so that even for a single D-brane one would expect to have a non-trivial potential that gives masses to ``off-diagonal string states".  One could afterwards impose a
Gauss' law constraint that ensures that the total charge on the D-brane is zero if the D-brane is compact. 
Thus we see that, on a first pass, the ABJM  theory looks completely different than a system of D-branes. 

The way to understand how the orbifold proceeds in this case is to study massive degrees of freedom on the moduli space of vacua. We find that a single membrane is not enough for this, as it has no massive modes.
Again, this is completely different than in a system of D-branes.
We consider then the next simplest case: a theory of two membranes. 

The scalar fields $\phi^A$ are now in bifundamental representations of $U(2)\times U(2)$, and
the moduli space is a symmetric product $Sym^2(\BC^4/\BZ_k)$, so it can be interpreted as the moduli space of two objects moving in $\BC^4/\BZ_k$.  These configurations can be described by $2\times 2$ diagonal matrices which (up to gauge equivalence) are of the form
\begin{equation}
\phi^A = \begin{pmatrix}
\phi^A_1&0\\
0& \phi^A_2
\end{pmatrix}\, .
\end{equation}
This requires a clarification. Since the scalars transform in the bifundamental representation, the matrices connect between two different vector spaces, so it is hard to interpret them as diagonal matrices on their own. Instead, 
we argue that they are diagonal if all the matrices $\phi^A \,\bar\phi_B$ commute with each other.\footnote{We denote with a bar the transposed complex conjugate matrix, $\bar{(\phi^A)^i_{i'}}\equiv (\phi^*_A)^{i'}_i$, reserving the dagger for the harmonic oscillator's creation operators that we shall introduce later. See the appendix for more details on our conventions.}
The coordinates of $\BC^4/\BZ_k$ can then be understood as eigenvalues of these matrices, similarly to what occurs for D-branes \cite{PolWit, BFSS}. 

\begin{figure}[tb]
\begin{center}
\psfrag{0}{$0$}
\psfrag{v}{$v$}
\includegraphics[scale=.5]{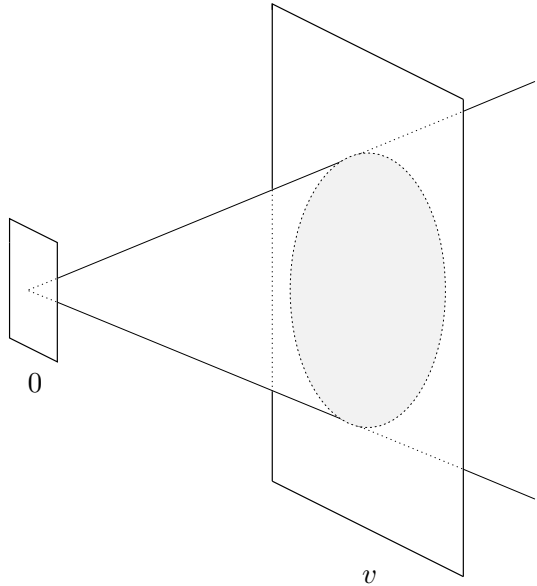} 
\caption{A configuration of two parallel M2-branes. The transverse space to the membranes is a real cone. One membrane is fixed at the tip of the cone while the other is moved to a distance $R=\ell_P^{3/2}v$ away. The off-diagonal modes stretching between the two membranes are called {\it membrane bits}, and wrap a preferred circle that we interpret as the M-theory circle. The tension of these bits scales like $\ell_P^{-3}$, suggesting that they are indeed two-dimensional objects, rather than strings. Looking at the scalar potential of the ABJM theory, one can also show that this tension is proportional to $1/k$, indicating that the radius of the M-theory circle gets shrunk by the orbifold.  In the string limit, in which this circle disappears, the membrane bits become string bits connecting two D2-branes. }
\label{cone}
\end{center}
\end{figure}

We can fix one of the branes at the origin, $\left|\left< \phi^A_1\right>\right|=0$, and move the other one to $\left|\left< \phi^A_2\right>\right|\equiv v$ (see fig. \ref{cone}).
This allows to define a field theory radius from the origin for the brane that we moved, given by $R \sim v$. Now, notice that $\phi^A$ has canonical dimension $1/2$, whereas a proper distance would 
have canonical dimension $-1$. To compensate for units, we find that
\begin{equation}
R = \ell_{P}^{3/2} v\, ,
\end{equation}
where $\ell_P$ will play the role of the Planck scale. For a membrane theory, the rescaled limit $v
\to 0$ in Planck units corresponds to the infrared (low energy limit), and this is the limit where the field theory conformal fixed point is supposed to live.

Now, the masses $m_{od}$ of the off-diagonal modes connecting the brane at $v$ to the brane at the origin, which we will call  {\it membrane bits}, are determined by the scalar potential (see Section \ref{masses-sec} for a precise computation). This is sextic by dimensional reasons  and is proportional to $1/k^2$ \cite{ABJM}. From this observation it follows immediately that
\begin{equation}
m_{od} \sim \frac{1}{k}v^2 = \frac{1}{k}\ell_P^{-3} R^2\,.
\label{tension}
\end{equation}
The factor of $\ell_P^{-3}$ should be interpreted in M-theory as the tension of an extended, two-dimensional object, and dimensional analysis tells us that the corresponding off-diagonal mode should be an M2-brane. 
This conclusion is independent of the details of the ABJM  model. Any conformal field theory in $2+1$ dimensions with canonical kinetic term and perturbative power counting would lead to the same
results. 

Notice that the scaling with the radius is proportional to $R^2$, and not to $R \ell_P$. This indicates that the membrane is wrapping an extended circle that scales with $R$. A theory that has this property suggestively requires a special circle where the membrane states prefer to wrap.
Such a special circle should be privileged from the field theory point of view. For example, translations around this circle, as far as the membrane bits are concerned, do not affect the membrane shape or tension as they happen along its world-volume. Therefore, for the off-diagonal degrees of freedom, the overall position of the ends on the special circle should not matter (after all, the brane is wrapped on it). We would think that this position is understood as a gauge artifact: there should be some gauge freedom associated with translations on this special circle.

Indeed, in the ABJM model, one finds that the full structure of moduli space is $\BC^4/\BZ_k$ only if one takes into account the gauge field degrees of freedom and the non-invariance of the Chern-Simons action \cite{DMPV}. A naive analysis based on gauge invariant scalar field configurations would suggest that the moduli space is a real cone over $\CP^3$ instead. Thus, in models where one expects a preferred circle, it should  be no surprise that the full set of isometries of the gravity dual might have different origins.  In particular, for the theories where one expects 
an $AdS_4\times S^7$ dual, the $SO(8)$ symmetry would have to be broken explicitly in the Lagrangian, and one would only be able to recover it from understanding the full quantum mechanical system, rather than some semiclassical approximation to it. This is what happens in the ABJM theory, where one only has an explicit $SU(4)$ symmetry, and the $SO(8)$ can only be recovered in the full quantum theory for $k=1$ and $k=2$.

Notice also that, since the tension in (\ref{tension}) is reduced by a factor $1/k$ relative to the theory without orbifold, one concludes that the membrane is wrapping a circle that gets cut into $k$ pieces in the process.

Now that we have made these various observations, we can finally address why the procedure of orbifolding along certain directions reduces to just changing the level of the Chern-Simons coupling. 

The basic idea is to consider the  M-theory geometry associated to the setup we are trying to describe. Consider a system of parallel M2-branes on flat space, and let us fix an origin on the transverse directions. This is an origin of $\BC^4$. We will choose to write the $\BC^4$
in spherical coordinates
\begin{equation}
ds^2 = dr^2 + r^2 d\Omega_7^2\,.
\end{equation}
We will also write the $S^7$ geometry as a Hopf fibration over $\CP^3$
\begin{equation}
\begin{matrix}
S^1 &\rightarrow & S^7\\
&& \downarrow\\
&& \CP^3\, .
\end{matrix}
\end{equation}

A similar structure can be obtained if one replaces the $S^7$ with any Sasaki-Einstein  manifold. We 
would have a $U(1)$ fibration over a complex manifold base, whose metric has constant curvature. In this way, the intuition can be applied to other field theories.

We will decree by fiat that the off-diagonal degrees of freedom connecting any two membranes away from the origin are membrane bits that wrap the $U(1)$ fiber completely. At fixed radius, the fiber is of constant size along the base.

\begin{figure}[tb]
\begin{center}
\psfrag{a}{M2}
\psfrag{b}{M2}
\psfrag{1}{$1$}
\psfrag{2}{$2$}
\psfrag{3}{$3$}
\psfrag{4}{$\cdot$}
\psfrag{5}{$\cdot$}
\psfrag{6}{$\cdot$}
\psfrag{k}{$k$}
\includegraphics[scale=.5]{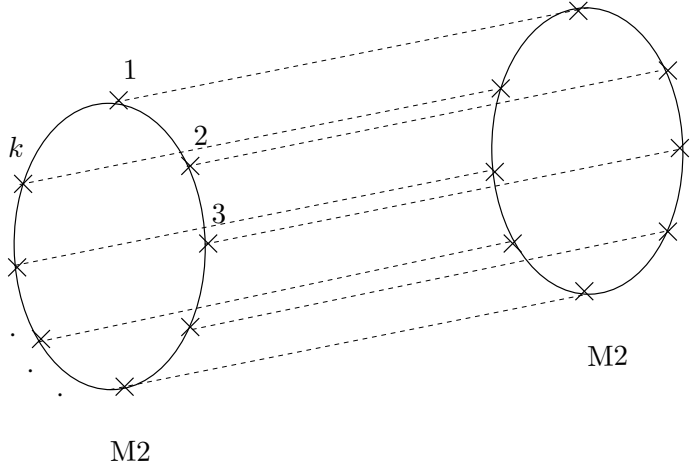} 
\caption{The membrane bit stretching between two M2-branes wraps completely the $S^1$ fiber over $\mathbb{C}P^3$, that we interpret as the M-theory circle. The orbifold procedure cuts this circle into $k$ segments and shrinks its size by a factor of $1/k$, thus changing the string coupling constant.}
\label{orbifold}
\end{center}
\end{figure}

 Now let us consider an orbifold where we orbifold by translations on the fiber, and shrink the effective size of the fiber by $k$. As far as the extended branes are concerned, the branes are point-like along the circle fiber, and one expects to see $k$ preimages of a brane in the 
orbifold (see fig. \ref{orbifold}). The positions on the base manifold are given by $r$ and a set of three complex coordinates characterizing a position on $\CP^3$, plus an angle on the fiber (let us call all of these coordinates $x$). The preimages differ only on the angle. 

On the other hand,  the membrane bits wrap the circle fiber and on the base they give a path between $x$ and $x'$. It should be the case that the membrane bits are those whose 
mass is lowest given the constraints of wrapping the fiber. 

In the orbifold, the new membrane bits are wrapping a circle which is $1/k$ times the size of the original circle. Thus, the mass of the new bits is given by 
\begin{equation}
m_{new} =\frac{ m_{cover} }{k}\,.
\end{equation}

Moreover, since the membrane bits touch all the preimages of the extended branes, they cannot be associated to any one such preimage. They seem to be automatically invariant 
under the group $\BZ_k$. For very small separation $\Delta$ between $x$ and $ x'$ on the base, 
these look like two membranes that are compactified on $S^1$. The off-diagonal modes wrap the fiber and we can interpret them in terms of  a reduction to type IIA string theory. The membrane bits can  be interpreted as  strings connecting between two D2-branes. Changing the
size of the fiber by a factor of $1/k$ only changes the string coupling constant, and reduces the string tension by a factor of $1/k$ relative to eleven-dimensional units, while it leaves the geometry of the base intact. Thus, locally, the Lagrangian describing the string motion should not change, except for the tension of the string. Similarly, the effective action for the strings will be the same as before, but with a modified string coupling constant.

All we need to make sure of is that for small separation we can approximate the off-diagonal masses by $\ell_s^{-2} \Delta$. This is a calculation that can be done readily in the ABJM theory, and we find that indeed it works that way. This also guarantees that in the D-brane limit the action looks like a commutator squared for the separation between the branes. 

We see that the main effect of the orbifold is to change the masses of the off-diagonal degrees of freedom by $1/k$ in Planck units. To calculate the effective string coupling constant dependence on $k$, we find that at radius $R$  
\begin{equation}
T_{D0}= \frac k{ R}\,,
\end{equation}
while the string length is given by
\begin{equation}
T_{string} = \frac1{2\pi \ell^2_s} = \frac R{2\pi \ell_P^3 k}\,.
\end{equation}
This translates in the scaling
\begin{equation}
\ell_s \sim \frac{\ell_P^{3/2} k^{1/2}}{R^{1/2}}\,.
\end{equation}
Now, the D0-brane tension in units of $\ell_s$ scales like $1/g_s$, and we find that
\begin{equation}
g_s^{-1}\sim T_{D0} \,\ell_s \sim \frac k{ R}\frac{\ell_P^{3/2} k^{1/2}}{R^{1/2}}= {k^{3/2}}\frac{\ell_P^{3/2}}{R^{3/2}}\,,
\end{equation}
so that keeping $\ell_P$ and $ R$ fixed while taking $k$ large is the limit of weak string coupling. Taking $R$ large makes the system go to strong coupling.

We should also mention the Chern-Simons part of the Lagrangian. Since this part does not contain any local degrees of freedom on its own, it only plays a role when 
we try to compute the moduli space, when we shall see that the equations of motion of the gauge field give rise to a modified Gauss' law constraint. It also plays a role in the beta functions for the coupling constant. Consistency with supersymmetry requires that we multiply the level of the Chern-Simons action by $k$. This guarantees conformal invariance, and the correct moduli space computation. We will see later on that this has other physical implications
that are required in order to have a match in the AdS/CFT correspondence. The quantization of $k$ is required for quantum consistency, but it cannot be proved at the classical level.

Finally, we should mention that it is possible to study the membrane dynamics by putting two different membranes at some non-trivial relative velocity. We can then ask if one can recover the corresponding gravitational potential between them by integrating out the off-diagonal degrees of freedom, {\it \`a la} M(atrix) theory \cite{BFSS}. One should be able to setup calculations similarly to \cite{Berenstein:1997vm},
and then one should produce a velocity dependent effective potential. The conformal scaling arguments 
are similar to \cite{Verlinde:2008di}, so that when one integrates the off-diagonal modes, one gets an effective action that goes as 
\begin{equation}
S_{Eff} \sim \int d^3 x \frac{|\Delta\dot\phi|^2}{|\Delta \phi|^6}\,.
\end{equation}

From here one should be able to compare with M-theory, where the interaction between the membranes has the same qualitative form, but where one measures the effective potential between the branes with the eleven-dimensional gravitational constant. The corresponding Green's function also decays like $1/r^6$ (there are eight transverse coordinates).\footnote{This calculation in conformal ${\cal N}=4 $ SYM shows a separation dependence  that decays as $r^{-4}$, which matches the type IIB setup, where  we get a Green's function for a Laplacian in six transverse coordinates to a D3-brane}

 To do the field theory calculation one needs the exact mass of the off-diagonal modes. We will compute this later in the paper (see Section \ref{masses-sec}). Notice that if we make stacks of branes (let us say of $S_1,S_2$ such branes), the one loop contribution already gives a term proportional to the product $S_1 S_2$ from summing over colors of the unbroken gauge group. This works similar to the case of D-branes. Also, in the case where the M2-branes are close to each other and can be interpreted locally in terms of type II backgrounds, we should get a result that is compatible with the type IIA setup, which would look like 
\begin{equation}
S_{Eff} \sim \int d^3 x\, G_{10} \frac{|\Delta\dot\phi|^2}{|\Delta r|^5}\,,
\end{equation}
where one understands that $G_{10}$ depends on position and $r$ is the distance between the D-branes in the type II theory. A complete analysis of this one loop calculation is beyond the scope of the present paper. In this vein, it should also be interesting to dimensionally reduce the system to 
a $1+1$ and $0+1$ field theory and to try to compare with the $\BC^4/\BZ_k$ orbifolds of type IIA and type IIB string theory, but where the M-theory circle and torus is taken along the transverse directions to the orbifold, rather than along the fibre of the Hopf fibration. This might in the end provide an alternative formulation of M(atrix) theory for type IIB string theory that is different from the D-instanton matrix model \cite{IKKT}.


\subsubsection*{Wilson loops}

Before moving on to the analysis of the field theory on $S^2\times \mathbb{R}$, we discuss briefly Wilson loops in the ABJM theory, emphasizing the differences with respect to the ${\cal N}=4$ SYM case. We will come back to this point later on in Section \ref{masses-sec}, where we will give a more precise definition of these operators. 

In the absence of matter, the Wilson loops in Chern-Simons theory compute topological objects as knot invariants \cite{Wittencs}, and are somewhat less interesting than in four dimensions, where they can be used as an order parameter for confinement. For theories coupled to matter, on the other hand, we expect to find a similar structure to the four-dimensional case of ${\cal N}=4 $ SYM, where the definition of these operators involves the scalar fields in a non-trivial way \cite{Rey,Malda}.

The idea is similar to the one described in \cite{Malda}: an off-diagonal membrane bit that is wrapped on the special circle  can connect between two membranes. The mass of the membrane bit can be seen as the difference of the areas of two cones:
\begin{equation}
m\sim T( R^2-r^2) \label{eq:massdiffcone}
\end{equation}
where $T$ is the tension of the membrane bit,  $R$ is the distance from the origin of a probe brane, and $r$ is the expectation value of some scalar field in the in the ${\cal N}=6 $ theory. 

Since $r^2$ is a matrix that can be in the adjoint of the group, we have that the corresponding Wilson loop observables should have the following  schematic form
\begin{equation}
W =\frac{1}{N} \tr \,\mbox{P}\left(\exp \oint i A  + \beta (\phi^2) ds\right)
\end{equation}
The part that is proportional to $\beta(\phi^2)$ is a correction to the mass of the particle along the given trajectory. Notice that this is positive definite. The bare mass that is proportional to $R^2$ has been eliminated from the observable, because it contributes trivially as the perimeter of the Wilson loop. This definition differs from the ${\cal N}=4$ SYM one, whose exponential is linear in the scalar fields \cite{Malda}. 

We should notice that the fields $\phi$ have canonical dimension $1/2$, so the Wilson loop described above is dimensionally correct and therefore $\beta$ is just a  number. This means that the observable above is well defined as an observable in a conformal field theory setting, where one has no additional scale in the problem than the geometry of the Wilson loop itself. As such, the expectation value will be independent of size, although it will in general depend on the shape. Also, notice that in principle the orientation of $\phi^2$ can depend on how one moves along the loop.

 This Wilson loop observable can be calculated from the mass formula for the off-diagonal membrane bits. We will see later on  in Section \ref{masses-sec} that the masses for off-diagonal bits can take exactly the form suggested in (\ref{eq:massdiffcone}). One can then calculate the expectation value of these Wilson loop observables perturbatively, but this is beyond the scope of the present paper.
Our mass calculation for off-diagonal modes can be used as input for these types of observables. In ${\cal N}=4 $ this mass is just the distance between two D-branes. In this case, the mass calculation is harder, but we will show that it has the required form and that indeed it gives the difference of two cones for some particular orientation.

In terms of the field theory degrees of freedom as bifundamental fields, the notion of $\phi^2$ will be given by $\phi^A \bar\phi_A$ for some fixed $SU(4)$ label $A$ of the R-symmetry. This is a Hermitian (positive) matrix in the adjoint of $U(N)$. This is basically a vector pointing in the direction of the brane, but with the phase of the location of the brane in the M-theory circle removed. 


\section{Dimensional Reduction to spherically symmetric configurations}
\label{sec:dimred}

One of the most useful aspects of conformal field theories is the existence of the operator/state correspondence. This is a correspondence between local operators inserted at the origin in Euclidean space, and states of the quantum field theory compactified on a sphere times time.

The essence of the correspondence can be understood in terms of classical physics. 
The idea is that we can write the Euclidean flat space in $D+1$ dimensions in radial coordinates as
\begin{equation}
ds^2 = dr ^2 +r^2 d\Omega_D^2\, ,
\end{equation}
and then we can do a local conformal rescaling of the metric by a factor of $1/r^2$, to obtain the rescaled metric
\begin{equation}
\tilde{ds}^2= d\tau^2 + d\Omega_D^2\,,
\end{equation}
where we have introduced a new variable $\tau\equiv  \log r$.

 These local rescalings give us a way to relate the conformal field theory on two different backgrounds if we impose that the theory is Weyl invariant, {\it i.e.}, it only couples to the conformal class of the metric, but not to the metric itself. Now, any field in the theory will transform with some weight under the Weyl rescaling, so that
\begin{equation}
\tilde \phi(x) = r^\Delta\phi(x) = e^{\Delta \tau} \phi(x)\, ,
\end{equation}
where $\Delta$ is going to be the dimension of the corresponding field.

If we now analytically continue $\tau$ to define a new time variable $i \tau\equiv   t$, we have 
that the Euclidean theory on $\BR^{D+1}$ with the origin removed can be related to the 
field theory on $S^D\times\BR$, where we have a Lorentzian signature.
In this system, we have the identification of the following vector fields
\begin{equation}
r\partial_r = \partial_\tau = i\partial_t\,,
\end{equation}
so that radial rescalings are equivalent to translations in $\tau$. Also notice that $\tau\to -\infty$ is the origin of the system.

Now consider solving the differential equations associated to the Euclidean system, with an operator insertion at the origin (thinking of this as a classical source for the fields, where the field equations are not satisfied). Since in the new coordinate system the origin is removed, we should replace the notion that there is a source at the origin by altering the boundary conditions of the fields at $\tau\to -\infty$. When we do this in the Lorentzian theory, we have a boundary
condition in the infinite past. This is the same thing as stating that we have a well prescribed 
initial condition for the fields in time $t$ for some slice $t_0$, where $t_0$ can be taken arbitrarily far into the past. To connect operators and initial data, all we need to do is solve the 
classical equations with prescribed boundary conditions and cut it at $\tau=0$. Then we can do the analytic continuation (Wick rotation) of the field configuration and its derivatives at the given time. 

Let us now consider passing to the quantum theory. If we insert an operator at the origin, 
this will correspond to some boundary condition at $\tau\to -\infty$.
We should also be able to understand such a boundary condition as a boundary condition on
the Lorentzian infinite past. These initial conditions
in quantum mechanics are nothing else than states in the quantum field theory. With
the time evolution operator the choice of where we cut the state open at different times can be related to each other by simple time evolution of the system.

The basics of this correspondence can be understood by the following data. Given 
some operator at the origin, $\CO_\alpha(0)$, there is a corresponding state $\ket{\alpha}$ associated to it.
A vector field associated to some symmetry in quantum mechanics gets realized by an 
operator. The operator associated to time evolution is the Hamiltonian. 
Thus, for a state $\alpha$  that is an eigenstate of the Hamiltonian  we have
\begin{equation}
H\ket \alpha = \alpha \ket\alpha\,.
\end{equation}
In the Euclidean theory the operator that generates motion along $r\partial_r$ is the dilatation operator $\Delta$. The corresponding evolution equation is a commutator relation (like the Heisenberg equation for operators in quantum mechanics), so we have the corresponding diagonalization process of $\Delta$ on operators:
\begin{equation}
i[\Delta, \CO_\alpha(0)] = \Delta_\alpha \CO_{\alpha}(0)= \alpha \CO_{\alpha}(0)\,.
\end{equation}
This is, the energy of a state gets identified with the scaling dimension of the associated operator. 

The isomorphism at the quantum level between states and operators can be easily established in free field theories.
However, we are analyzing field theories in three dimensions that are not free, but that in some regimes can be understood in terms of a perturbative expansion around some free degrees of freedom.  If we turn on interactions, the operator problem that one needs to understand is the generation of anomalous dimensions for operators, while in the state setup, one is doing 
ordinary time independent perturbation theory. 

A particularly interesting set of operators that one can analyze are those that belong to the chiral ring. These are operators whose free field dimension (if such a description exists) are not renormalized by varying the coupling constant. 
However, we can have situations where even the chiral ring has some operators that cannot be described in terms of free fields, and as such, it is hard to imagine that conventional perturbation theory will be useful to describe such objects.

This is the case where a detailed understanding of the dynamical system of the field theory compactified on $S^2$ can lead to insight into the corresponding degrees of freedom. We will thus analyze the reduction of degrees of freedom on a sphere. We will be interested mainly in spherically invariant configurations (these are the ones that describe scalar operators and can acquire vevs in the Euclidean theory when we spontaneously break conformal invariance, but not Lorentz invariance).  Once given such a spherically invariant configuration, one can then do a spherical harmonic decomposition of fluctuations and analyze nearby states to a given state. Since all of the matter degrees
of freedom that we are analyzing  in the ${\cal N}=6$ field theories in three dimensions carry charge under some gauge group degrees of freedom, we will analyze first the 
spherically invariant field configurations that are allowed for a $U(N)\times U(N)$ connection on $S^2\times \BR$. Then we will analyze the bosons and we will only make passing remarks 
about fermions, as these do not get vevs in the classical theory. 


\subsection{Gauge fields}

We are considering a $U(N)\times U(N)$ gauge field on $S^2\times \BR$. 
In order for the gauge field to be spherically symmetric, we need the following conditions (we will analyze a single $U(N)$ first).

First, it must be the case that the gauge field strength is covariantly constant on $S^2$. This is the natural  notion of spherical symmetry that is gauge invariant. The gauge field strength, if we use standard coordinates $\theta,\varphi$ on the $S^2$, consists of the three adjoint components $F_{\theta\varphi}$, $F_{0 \theta}$ and $F_{0 \varphi}$. The first one is a two-form on $S^2$, while the other two give together a one-form on $S^2$.

Since we require that the gauge field is covariantly constant on $S^2$, we get that $F_{\theta\varphi}$
is covariantly constant on its own. We can write $F_{\theta\varphi}\, d\theta\,  d\varphi= \Phi(\theta,\varphi) \, d ( vol_{S^2})$, where $\Phi$ is a covariantly constant scalar field in the adjoint of $U(N)$, since the volume form is spherically symmetric. We can use gauge transformations to diagonalize $\Phi$, and the covariantly constant constraint tells us that the eigenvalues of $\Phi$, which we call $\Phi^i$, are constant on the sphere. Moreover, these configurations for the gauge connection $A_\theta, A_\varphi$ are equivalent to those that satisfy the two-dimensional Yang-Mills equations. These have been analyzed in detail by Atiyah and Bott \cite{Atiyah:1982fa} over arbitrary Riemann surfaces. The eigenvalues of $\Phi$  are to be thought of as fluxes for $U(1)^N$, which is the unbroken gauge group associated to such a scalar field configuration. 
These fluxes are quantized and therefore cannot change in time continuously. 

Also, if the electric field strength is covariantly constant, then we have that 
\begin{equation}
D\wedge (F_{0i} d\sigma^i)=0\, , \qquad (i=\theta,\varphi)\, .
\end{equation}
Together with the Bianchi identity we find that
\begin{equation}
D_0 \Phi =0\,,
\end{equation}
so if we choose a gauge where $\Phi$ is diagonal for all times, we find that $[A_0,\Phi]=0$, so that
a non trivial time-dependent gauge potential has to satisfy $A_0\in U(1)^N$. Also, we get the 
constraints $[A_\theta,\Phi]=[A_\varphi,\Phi]=0$ (these are the off-diagonal equations associated to a covariantly constant configuration in the gauge where $\Phi$ is diagonal). One finds very quickly that in this gauge $A_\theta$ and $A_\varphi$ should be time independent. 

Thus we have reduced everything to $U(1)^N$. The gauge field strength $F_{0i}= \dot A_i$ for each eigenvalue determines a vector field in $S^2$ that should be covariantly constant. Such vector fields do not exist (the two-sphere cannot be combed).
If one has degeneracies of the fluxes, one can use a more sofisticated argument, but one ends up with essentially the same answer.\footnote{A more topological argument is that the families of solutions of Yang-Mills equations are parametrized by representations of a central extension of the fundamental group of the Riemann surface \cite{Atiyah:1982fa} (these are holonomies for fixed paths), and that the sphere has no fundamental group, so these are trivial.} 

In summary, the allowed spherically invariant configurations for the gauge field are almost trivial. They are completely characterized by constant fluxes in the Cartan basis, and they break the gauge group 
to $U(1)^N$ if the fluxes are all different. If we consider  $U(N)\times U(N)$ gauge fields, the result is the same, so we have fluxes $\Phi^i$ for the first $U(N)$ and fluxes $\tilde\Phi^{i'}$ for the second one. Also, the fluxes specify a unique connection up to gauge transformations.
These connections are connections on direct sums of line bundles.


\subsection{Scalar fields}
\label{scalars-sec}

As we described above, the spherically invariant dimensional reduction of a gauge field on $S^2$
is not completely trivial. We will need a gauge field that can be reduced to $U(1)^N\times U(1)^N$. This will be characterized by fluxes $\Phi^i$ and $\tilde \Phi^{i'}$. We analyze now the bifundamental complex scalar fields $\phi^A$  which are charged under these groups.

To find spherically invariant configurations, we fist decompose the $\phi^A$'s according to their gauge indices with respect to the labels $i,i'$. The fields $(\phi^A)^i_{i'}$ experience a magnetic monopole background on the sphere of strength $\Phi^i-\tilde\Phi^{i'}$, and therefore their spherical harmonics will be given by monopole spherical harmonics. They will carry angular momentum given by $\ell+|m_i-\tilde m_{i'}|/2$, for $\ell=0, \dots, \infty$ \cite{MSH}. These can be spherically symmetric only 
if $\ell=0$ and $ m_i=\tilde m_{i'}$. Thus, the possibility of non-trivial scalar vevs forces us to consider 
configurations where the $U(1)^N\times U(1)^N$ unbroken symmetry gets reduced to a diagonal 
$U(1)^N$, where we have identified the $i$ labels and the $i'$ labels. This is the situation where there are no degeneracies between the fluxes. With these degeneracies one can have a slightly more intricate structure for the scalar fields.

Given a Chern-Simons coupling of the gauge fields, and standard kinetic terms for the scalars $\phi^A$, we find that the equation of motion of $A_0$ reads
\begin{equation}
k\Phi^i = (\phi^A \dot{ \bar\phi}_A-\dot \phi^A \bar \phi_A)^i_i= Q_i\label{eq:quant}\,,
\end{equation}
where $k$ is the level of the Chern-Simons theory and
where $Q$ is the charge carried by the fields. Similarly for $\tilde \phi$, for which we get $\tilde Q_i=k\tilde \Phi^i$.

This is, the gauge constraint tells us that the ``angular momentum" along the charge direction is quantized already at the classical level. It is also only carried along the Cartan directions. This condition is a generalization of the analysis of monopole operators in Chern-Simons theory done in \cite{BKW}.

Finally, we should look at the induced dynamics of these free scalar configurations. We find that besides the usual kinetic term that is induced by the dimensional reduction, and which is given by  $\tr(\dot\phi^A\dot{ \bar \phi}_A ) $, we also need a mass term. 
This mass term arises because we are considering a conformal field theory compactified on $S^2\times\BR$ background metric, and the conformal symmetry requires a conformal coupling to the background curvature. A detailed derivation of this coupling can be found in \cite{Wald}.

For the scalar fields, if they have canonical kinetic term, then their conformal coupling gives a mass equal to $1/4$. This is in conventions where the time units on $S^2\times \BR$ are chosen such that the eigenvalues of the Hamiltonian correspond directly to the spectrum of dimensions of local operator insertions (see the appendix for more details). After all, $S^2\times \BR$ is related by a conformal rescaling to radial quantization, where the dilatation operator gets transformed into the Hamiltonian in $\BR$. The mass can be checked by realizing that the free field operator $\phi^A$ is equal to a single excitation of the s-wave harmonic of $\phi^A$ on $S^2$. Since the dimension of $\phi^A$ is $1/2$ and this should be equal to the mass, we find that $m^2=1/4$.

We also find immediately that the $Y_{\ell m}$ spherical harmonics of $\phi^A$ would have a mass squared 
equal to
\begin{equation}
m^2_{\ell m} = \ell (\ell +1)+ 1/4= (\ell +1/2)^2
\end{equation}
This translates into a frequency of oscillators given by $w_\ell =\ell +1/2$.
This also exactly matches the dimensions of descendants of $\phi^A$, given by $\partial^\ell \phi^A$.
This is how the operator/state correspondence works for free fields: the operator $\phi^A$ translates to the creation operator of the s-wave quanta of $\phi^A$, and the descendants of $\phi^A$, given by $\partial^{[\ell]}\phi^A$, translate to the creation operators for $\phi^A$ with
angular momentum equal to $\ell$. We see that the energy of the oscillators matches the dimensions of the corresponding fields.

In the case where there is a monopole background with flux $2s$, the corresponding mass squared is 
modified to
\begin{equation}
m^2= (\ell+1/2)^2-s^2
\end{equation}
for $\ell= s , s+1,\dots ,\infty$ (see also \cite{BKW} for the spectrum of the associated Dirac equation). The quantum number $\ell$ is the total angular momentum of the excitation. Both $\ell$ and $s$ can be half integers.  Notice that for $s\neq 0$, the square root is generically non-rational. This means that in the perturbative limit, for these monopole operators, we find that their excitations we do not have rational conformal weights. This implies that these ``excited'' operators cannot be thought of in terms of free fields. This is natural, as the constraint that we use to solve for the gauge field is non-linear. These solutions should be considered as non-perturbative saddles.

Finally, any scalar potential that the $\phi^A$ has should be kept in the dimensional reduction on the sphere.
In the case at hand of ${\cal N}=6$ gauge theories, for gauge configurations that break the group to $U(1)^{N}\times U(1)^N$ and where $\phi^A$ is spherically symmetric, the scalar potential
vanishes automatically: after all,  the only configurations allowed are diagonal, and these are the minima of the matter field potential. These are the moduli space of vacua. If one has degeneracies of the fluxes, then one would also want to understand the potential for the scalar fields in more detail. We will do this in future sections.

The main result we should take home is that for a field $\phi^A$ to get a non-trivial vev that is spherically symmetric, the fluxes under which the field is charged as a bifundamental have to be identical.

For fermions, the situation is not too different. The main difference is that fermions 
start as sections of $\CO(1)\oplus \CO(-1)$, and we have to tensor this representation by 
the vector bundle structure. One would have to consider monopole spinor harmonics if the fluxes don't match. The fermion excitations are not as interesting as they do not match the BPS bound that we are after and moreover in the classical theory their vevs are zero.


\section{Free field description of BPS local operators, giant gravitons, and their fluctuations}
\label{sec:ffBPS}

As we have discussed so far, the operator/state correspondence permits one to establish a one-to-one map between local operators in the Euclidean field theory and members of the Hilbert space of states of the quantum field theory on the Lorentzian $S^2\times \BR$.
The generators of momentum and special conformal transformations $P_\mu$ and $ K_\mu$ in the Euclidean theory become operators in the Hilbert space of states. The natural condition for unitarity on $S^2\times \BR$ is that these two operators are related so that $P$ is the adjoint of $K$, namely
\begin{equation}
P_\mu = (K_\mu)^\dagger\,.
\end{equation}
The $K_\mu$ operators lower the dimension of operators, while the $P_\mu$ increase them, since they obey the algebra $[\Delta,K_\mu]=i K_\mu$ and $[\Delta,P_\mu]=-i P_\mu$. The dilatation operator $\Delta$ becomes the Hamiltonian of the theory on $S^2\times \BR$, so $K_\mu$ lowers the energy and $P_\mu$ increases it.

The existence of a vacuum state with minimum energy implies that in any unitary representation of the conformal group there is a lowest weight state, a conformal primary, that is annihilated by all the $K_\mu$'s. 

In the free field regime, all states can be described in terms of a Fock space of states,  with an additional condition that guarantees gauge invariance of the configuration. In the ABJM model, the free field limit occurs whenever the Chern-Simons level is large, $k\to\infty$, at fixed $N$. This is a situation where one can expect a string description {\it \`a la} 't Hooft if one takes $N$ large with $N/k$ fixed. For small values of $N/k$, the theory can be analyzed perturbatively.

 If we take quantum fields on $S^2$, and we decompose the fields into spherical harmonics, we end up studying the problem that we addressed in the previous section. The individual oscillators with the lowest energy are s-waves of the scalar fields on $S^2$.
 At total occupation number $s$,  the states with the smallest energy will be of the form 
 \begin{equation}
 C^{A_1 \dots A_s} a_{A_1}^\dagger \dots a^\dagger_{A_s}\vac\,,
 \end{equation}
where the $a^\dagger$'s create the s-wave part of the scalars on the sphere.

A BPS inequality for these representations relates the R-charge of such a state to the dimension of the operator. The R-charge is additive in constituents. 
Thus, if the raising operators all have the same R-charge, the states described above will have the smallest energy given the R-charge, and they are the candidates for BPS states.

To belong to the Hilbert space of states, the above state has to satisfy the gauge constraints
of the theory. This makes $C^{A_1\ldots A_s}$ into a gauge invariant tensor under the group labels. If we look at the field content of the ABJM theory, one can write it in terms of an ${\cal N}=2$ superspace formulation (this is the familiar ${\cal N}=1$ superspace formulation of four-dimensional theories) \cite{Benna:2008zy}. 
With this convention the ABJM theory consists of four chiral multiplets $A_a=(A_1, A_2)$ and $B_{\dot a}=( B_1, B_2)$ (the same chiral field content as the Klebanov-Witten field  theory in four dimensions \cite{Klebanov:1998hh}). The fields $A_a$ and $ B_{\dot a}$ have conformal dimension $1/2$,  the $A_a$ transform in the $(N, \bar N)$ of $U(N)\times U(N)$, while the $B_{\dot a}$ transform as the $(\bar N, N)$. Moreover, there is an $SU(2)\times SU(2)$ symmetry of the superpotential, under which the $A_a$ transform as the $(\frac 12,0)$ and the $B_{\dot a}$ transform as $(0,\frac 12)$.  Both $A_a$ and $B_{\dot a}$ have the same R-charge.

The first guess for a simple BPS state would require us to maximize the R-charge, so it will be some combination of $A_a$ and $ B_{\dot a}$ raising operators. For simplicity we will  choose to maximize the $SU(2)\times SU(2)$ weight of this state, and one can generate other states by $SU(2)\times SU(2) $ rotations. In the ${\cal N}=6$ theory, these $SU(2)$ symmetries are part of the R-symmetry group, so maximizing the $SU(2)$ R-charges gets us to a highest weight state of the R-symmetry group compatible with the ${\cal N}=1$ splitting. Thus, we should build a state out of $A_1$ and $B_1$ oscillators only.

The state will have the qualitative form $(A_1)^{j_1}(B_1)^{j_2}$ (with gauge indices that we suppress for the time being). 
To cancel the $U(1)$ gauge charge of this state in the perturbative setup we need $j_1=j_2$.
Let us differentiate between the first and second gauge group by using the notation $U(N_1)\times U(N_2)$. The  charge carried by the $A$ states will be given by a tensor with $j_1$ upper indices of $U(N_1)$ and $j_1$ lower indices of $U(N_2)$. The roles of $N_1$ and $N_2$ are reversed for $B$. In order to get a gauge invariant object, we need to contract the upper and lower $N_1$ and $N_2$ indices (remember that for $U(N_1)$ the only invariant tensor is $\delta^{i}_{j}$). This implies that the $A$ and $B$ are contracted 
always in the combination $(A_1B_1)$, and if we look at this object as an $N_1\times N_1$ matrix, the indices are contracted by taking traces. Thus, a collection of states that will be gauge invariant are given by taking products of traces
\begin{equation}
\prod_{j} [\tr(Z^j)]^{n_j}\vac\, ,
\end{equation}
where $Z$ is  a raising operator, given by $Z\equiv(A^\dagger_1 B^\dagger_1)$.
This is very similar to the set of half-BPS states in ${\cal N}=4 $ SYM theory. If we remember that $Z$ is composite, this construction would be similar to the set of chiral ring states with maximum $SU(2)\times SU(2)$ charge in the Klebanov-Witten theory.
The set of such half-BPS operators in ${\cal N}=4 $ SYM can be described by free fermions  \cite{CJR,Btoy}, and it is convenient to think of these systems as an integer quantum Hall droplet. This set of states for the ABJM theory can be described in terms of a matrix quantum mechanics for rectangular matrices, which also admits a description by free fermions which is very similar (see \cite{Bqhd}). The isomorphism to free fermions proceeds by
writing the tensor structure of the upper $U(N_1)$ gauge indices in terms of a 
Young tableau. Since the $A$ are bosonic operators, the symmetry properties of the upper indices is mirrored in the lower indices. Thus the $U(N_2)$ gauge group tableau is the same as the $U(N_1)$ tableau. Once written in terms of the Young tableaux, the representation of $U(N_1)$ is irreducible.  In order for the $B$ indices to contract and give a gauge invariant, the tableau for the $B$ gauge indices needs to be the same as that for the $A$ indices. This is because in the tensor product of two irreducible representations of $U(N)$, $\Lambda_1\otimes \Lambda_2$, one can have the trivial (gauge invariant) representation only when $\Lambda_1= \Lambda_2^*$.
One can also show that the states associated to two different Young tableaux are orthogonal.

The individual fermions in the droplet are the eigenvalues of the matrix $Z$. There are $N(=N_1=N_2)$ such eigenvalues. The rows of the Young tableaux can be related to the excitations of the free fermions relative to the Fermi sea. Thus, states that are related to Young tableaux with a single long row correspond to giving a lot of energy to a single fermion (exciting an eigenvalue), while the operators that correspond to making totally antisymmetric representations correspond to exciting hole states in the Fermi sea. The individual traces are edge excitations of the quantum Hall droplet. Incidentally, the condition $N_1=N_2$ is not necessary for the free fermion picture. Some of the details change, in particular, the number of fermions is the smallest number between  $N_1$ and $ N_2$.

Part of the interest in describing these more complicated states is that in the case of ${\cal N}=4 $ SYM it is known that these particle and hole states in the quantum Hall effect become two different types of
giant gravitons in the dual gravitational theory, as described geometrically in \cite{HHI,GMT}. 

The ones that give a lot of energy to a single eigenvalue can be interpreted easily as a classical  configuration that has one large eigenvalue and all others equal to zero \cite{HHI} (see also \cite{Btoy}). In the field theory, this is a Higgs branch, as the gauge symmetry is spontaneously broken. These configurations can be thought of as matrices where 
\begin{equation}
Z= \diag(z_1\, e^{i t}, 0, \dots, 0)\,.
\end{equation}
The time dependence arises from solving the equation of motion for $Z$. 
In terms of the $A_1, \,B_1$ matrices, we have that
\begin{eqnarray}
A_1 \sim \diag (a_1\, e^{i t/2}, 0, \dots ,0)\\
B_1 \sim \diag(b_1\, e^{i t/2}, 0, \dots ,0)
\end{eqnarray}
To solve the classical Gauss' law constraint, we need to require that  $|a_1|= |b_1|$. 

The off-diagonal fluctuations connecting the giant to the other degrees of freedom become massive, and the effective theory of light fluctuations of the giant is in essence the field 
theory for a single M2-brane.

Thus, large eigenvalues should correspond to M2-brane giants in the $AdS$ dual. 
Because the gauge invariance constraint requires that the momentum of the M2 brane along the fiber of the Hopf fibration over $\CP^3$ describing $S^7$ vanishes exactly, the 
quantum M2-brane is in an eigenstate of the appropriate angular momentum and should be thought of as a delocalized brane in the fiber direction. This same reasoning applies to the gravitons.

The giant growing on the sphere is built from the electric-magnetic dual of the brane growing into $AdS$. Thus, the holes should be describing some M5-brane wrapping a particular submanifold of $S^7$. 

In the string limit, where $N,\, k\to \infty$ with $N/k$ fixed, the M2-brane should be replaced by a D2-brane
growing into $AdS_4$, and localized at a point on $\CP^3$ (the fiber information is replaced by the type IIA dilaton and the type IIA RR vector field background on $\CP^3$).  The rotation quantum numbers on $\CP^3$ determine a geodesic on $\CP^3$ that the D2-brane follows.
Similarly, the individual gravitons will follow this same geodesic, but they will be at the origin of 
$AdS$ space in global coordinates.

The dual giant should be made from the electric-magnetic dual of the D2-brane wrapping some cycle. This is a D4-brane wrapping some cycle. Now we can ask if it is possible to find evidence that these states describe D-branes. Abstractly, a D-brane is characterized by stating that strings end on them. Thus we need to find candidate states for strings ending on these objects.

To do that, we would like first to have a description of string states near the corresponding geodesic in $\CP^3$. This is easiest to do in a large momentum limit. The idea is that strings are described by traces, where the ordering of matrices inside the trace matters. If we have a very fast moving string, it will have a lot of momentum, and the simplest such string state is
\begin{equation}
\tr(A_1^\dagger B_1^\dagger)^J\vac\label{vac-spinchain}
\end{equation}
where $J$ can be taken arbitrarily large, but such that $J/N<<1$.

Other states can be obtained by replacing finitely many of the $A_1, B_1$ by impurities.
For example, one $A_1^\dagger$ can be replaced by $A_2^\dagger$ or $\bar B_2^\dagger$, and 
$B_1^\dagger$ can be replaced by $\bar A_2^\dagger$ or $B_2^\dagger$.\footnote{Recall that we reserve the dagger for creation operators and the bar for transposed complex conjugates.} If we pick one such state, we have for example
\begin{equation}
|A_2, J_1; B_2, J_1+J_2+1\rangle=\tr((A_1^\dagger B_1^\dagger)^{J_1} A_2^\dagger B_1^\dagger(A_1^\dagger B_1^\dagger)^{J_2} A_1^\dagger B_2^\dagger)\vac\,,
\end{equation}
where the $J$ labels indicate the position along the word where the impurity was inserted. 

In perturbation theory all of these states with $J_1+J_2$ fixed will mix, as they have the same classical energy and quantum numbers under the symmetry group. If we allow the positions of all defects to be arbitrary on the word, then the cyclic property of the trace identifies many of these words. This cyclic rotation can be thought of as a translation by one on the set of periodic words (this is a finite cyclic group). The invariance of the trace under permutations is the statement that the total momentum of the word vanishes (this momentum is valued in the dual group of the translations, and is therefore periodic -- this is usually called the quasi-momentum of the respective configuration). In the dual string world-sheet, this is the level matching constraint \cite{BMN}.
Fixed order planar diagrams can move the defects to the left or right by finite amounts, and therefore the Hamiltonian on this set of vectors can be described in terms of a spin chain with local interactions \cite{MZ}. In the large volume limit of the spin chain, large $J$, the defects are approximately decoupled. Approximate solutions for eigenstates of the Hamiltonian will be given by wave functions where each defect is described by an approximately free excitation. Each defect will have its own quasi-momentum, and there will be an associated dispersion relation (this is an effect of the translation invariance of the spin chain problem).

Thus, to leading order, the correct eigenstates of the Hamiltonian will be given by Fourier transforms of the basis found above with respect to the relative positions of the defects. Explicitly, for the impurities considered above, this reads
\bea
\ket{\psi}\sim \sum_{l=0}^J e^{2\pi i q l/J} |A_2,l;B_2,J+1\rangle\, .
\eea

For special spin chain Hamiltonians, the full problem is solvable. These are the integrable spin chains. The exact diagonalization can be performed in terms of a Bethe ansatz
around a preferred vacuum: the state $\tr(A_1^\dagger B^\dagger_1 A_1^\dagger B^\dagger_1\ldots)\vac$ that we described above in (\ref{vac-spinchain}). 
This problem was analyzed to two-loop order in the ABJM theory by Minahan and Zarembo \cite{Minahan:2008hf}, and they found that the spin chain was indeed integrable. For small quasi-momentum excitations, the system ends up being mapped in the dual theory to the plane wave limit around a null geodesic, which sits at the origin in $AdS$ space, and that is given by some orbit on $\CP^3$ determined by the quantum numbers of the state in question. For quasi-momentum of order $1$, the defects are called magnons in the spin chain setup. The dual description of these states are extended strings. The corresponding magnons in the sigma model are called giant magnon configurations \cite{Hofman:2006xt}. We will have more to say about these states later in Section \ref{disp-rel-sec}, where we will compute their dispersion relation to all orders in the coupling constant.

\begin{figure}[tb]
\begin{center}
\begin{tabular}{cc}
\psfrag{a}{$a$}
\psfrag{b}{$b$}
\psfrag{c}{$\dot b$}
\psfrag{d}{$\dot a$}
\includegraphics[scale=.25]{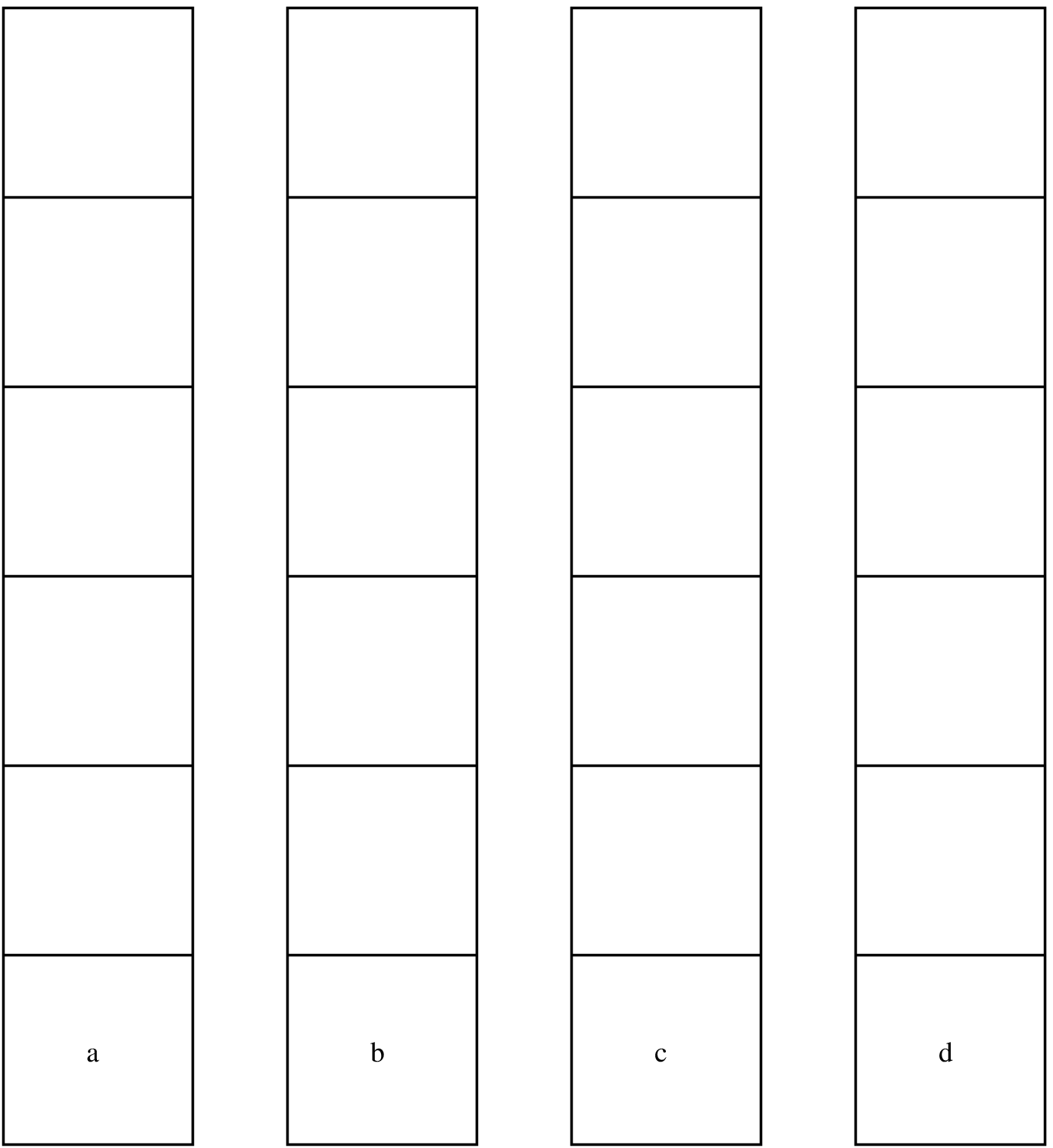}
&
\hskip 3cm 
\psfrag{a1}{$a$}
\psfrag{b1}{$a$}
\psfrag{c1}{$\dot b$}
\psfrag{d1}{$\dot b$}
\includegraphics[scale=.25]{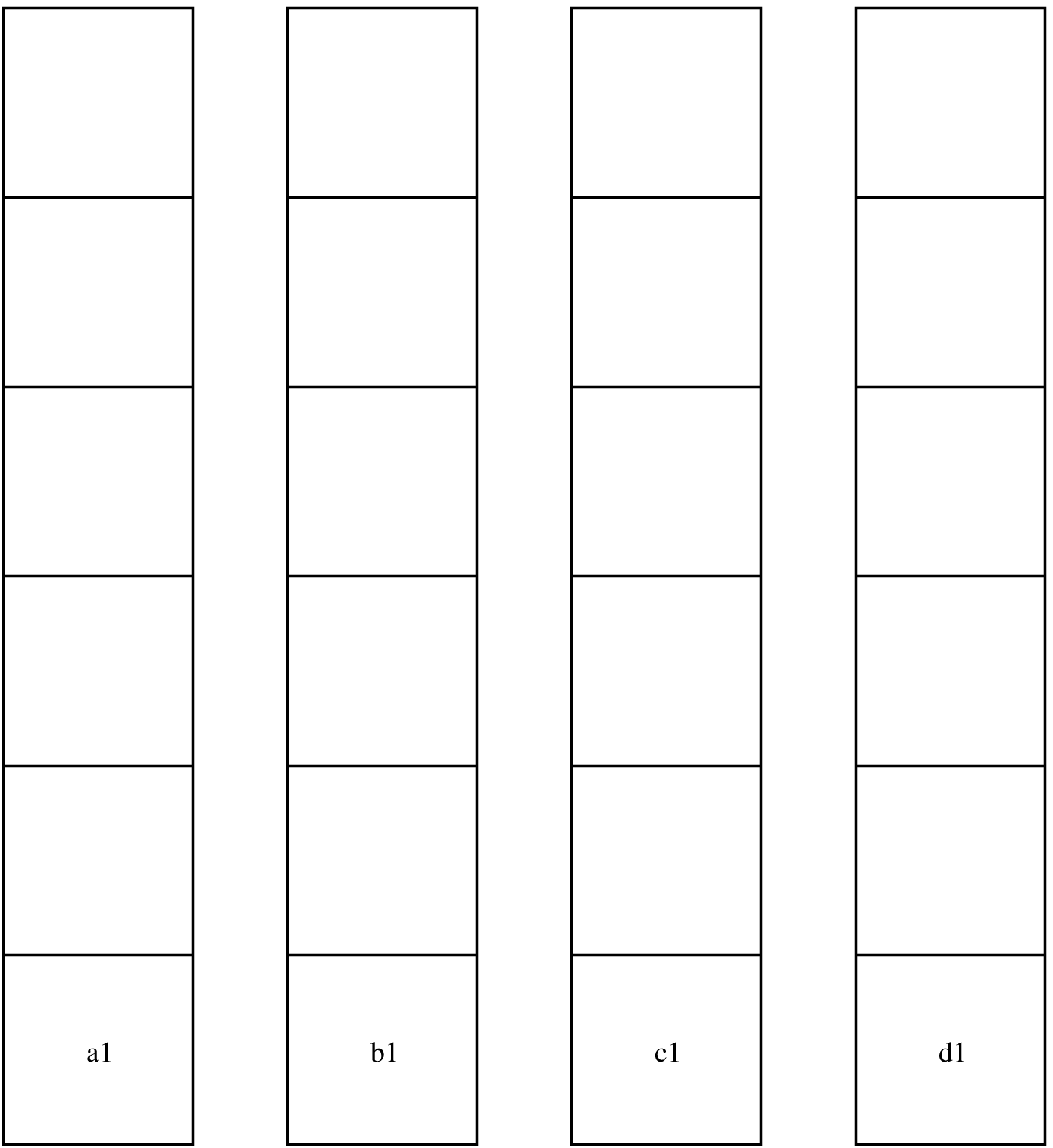}
\\
(a)&
\hskip 3cm  (b)
\\
&
\end{tabular}
\caption{Young tableaux describing the contraction of gauge theory indices for the giant gravitons in the ${\cal N}=6$ Chern-Simons theory. We consider for simplicity the case of maximal giants, {\it i.e.} operators in rank $N$ antisymmetric representation of the gauge group (in the figure $N=6$). Unlike ${\cal N}=4$ SYM, where one had just one upper and one lower index and one possible contraction thereof, we have now four indices and two possible contractions. The columns represent respectively $(A^a,\,\bar A_a,\, \bar B^{\dot a},\, B_{\dot a})$ and the boxes with letters inside are the marked boxes. We have then that (a) corresponds to the contraction mixing $A$ fields with $B$ fields while (b) correspond to the contraction of $A$ fields among themselves and $B$ fields among themselves.}
\label{young}
\end{center}
\end{figure}

If we cut a trace, the words have ends, and then one can envision having a spin chain dynamical system with a boundary.  Because now instead of a trace we have a matrix, there are floating gauge indices associated to the ends of a string. To interpret them in string theory, we would like to have these ends of the strings attached to D-branes. Thus, we should be able to tie the indices of words to gauge indices associated to D-brane configurations described by Young tableaux as above (see fig. \ref{young}).

A similar problem was solved for ${\cal N}=4 $ SYM in \cite{BBFH} (see also \cite{BHK,Berenstein:2003ah} for previous work in the subject). It was shown there that 
excitations of giants could be obtained by removing or attaching  marked boxes to Young tableaux. There was one such marking for each end of the string. These usually correspond to different types of indices (fundamentals and antifundamentals), so the marking was done in two  tableaux: the one for the upper indices, and the one for the lower indices. 
The decorations can always be placed on the 
corners of the tableaux.
The markings indicate the string ends. In simple setups one is adding a box to a row or column, and these are naturally string ends on the corresponding branes.

The corresponding spin chains with boundaries in ${\cal N}=4 $ SYM have been studied in \cite{BV, BCV1, BCV2, Correa:2006yu}. It was found there that these spin chains usually do not preserve the length of the spin chain \cite{BCV1} and that their spectrum of excitations is continuous in the large $N$ limit. This means that they cannot be immediately associated to systems that have a  Bethe ansatz description. The special case of a maximal giant did preserve the spin chain length. 
The study of these decorated tableaux has been carried further in \cite{ZA}. A similar analysis should be possible in this case.

In our case, for the maximal giant (corresponding to a Young tableau as in fig. \ref{young}) the state is given by
\begin{equation}
\det(A_1^\dagger B_1^\dagger)\vac = \det(A_1^\dagger) \det(B_1^
\dagger)\vac\label{eq:Dib1}
\end{equation}
and we notice that it factorizes. 
The operator 
\begin{equation}
\det(A_1^\dagger) \label{eq:Dib}
\end{equation}
 carries baryon charge with respect to two different baryon numbers, so (\ref{eq:Dib1}) is called a dibaryon.

\begin{figure}[tb]
\begin{center}
\psfrag{a}{D$4_a$}
\psfrag{b}{D$4_b$}
\psfrag{cp1}{$\mathbb{C}P^1$}
\psfrag{aa}{$\ket{aa}$}
\psfrag{bb}{$\ket{bb}$}
\psfrag{ab}{$\ket{ab}$}
\psfrag{ba}{$\ket{ba}$}
\includegraphics[scale=.5]{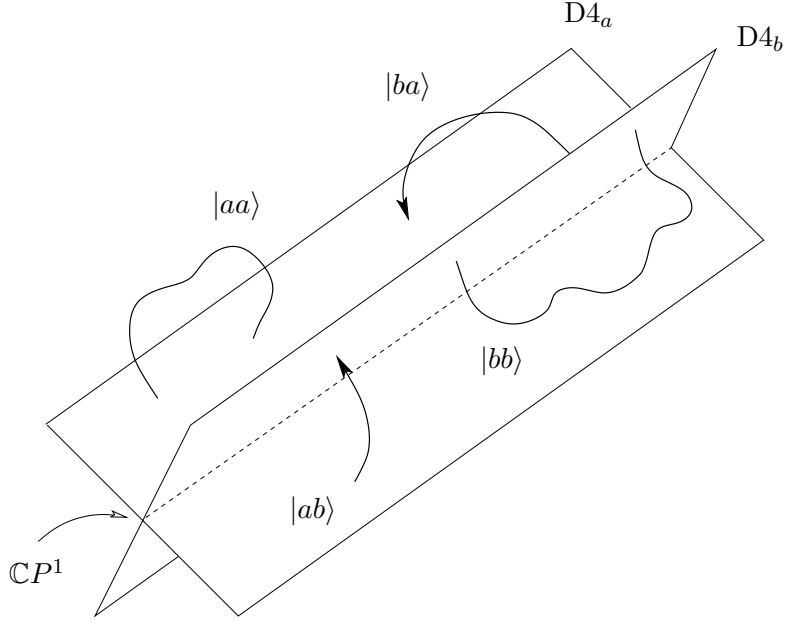} 
\caption{Intersecting D4-branes corresponding to a giant graviton. These intersect on a $\mathbb{C}P^1$ and wrap two different $\BC P^2$ submanifold in $\BC P^3$. There are four kind of strings stretching between them: the $\ket{aa}$ strings correspond to replacing a $A_1$ field with a composite word with the same index structure, and similarly the $\ket{bb}$ string corresponds to replacing one $B_1$. On the other hand, the $\ket{ab}$ and $\ket{ba}$ strings correspond to replacing both $A_1$ and $B_1$ fields at the same time.}
\label{branes}
\end{center}
\end{figure}

In the Klebanov-Witten theory each such dybarion corresponds to an individual D-brane. 
Thus, for the maximal giant we expect that (\ref{eq:Dib1}) degenerates into two D4-branes in $\CP^3$, each wrapping a different $\CP^2$ submanifold (see fig. \ref{branes}).
Let us call them D4$_a$ and D4$_ b$. These intersect on a $\CP^1$. We see that there are four classes of strings: $\ket{aa}$, $\ket{ab}$, $\ket{ba}$, and $\ket{bb}$ strings. These are further restricted to have vanishing string charge on them (this is a consequence of the Gauss' law on the D-branes themselves, which are compact).

These strings are easy to be related to excitations of  the determinant operator. For the $\ket{aa}$ strings  replace one $A_1$, by a composite word with the same index structure, and no $A_1$ on the beginning or end \cite{BHK}: such operators can be factorized into a trace times the dybaryon.
These boundary restrictions play the role of the Dirichlet boundary conditions on the open spin chain.
Similarly, for the $\ket{bb}$ strings we replace a $B_1$. For the $\ket{ab}$ strings, we need 
another string $\ket{ba}$ in the opposite direction. These are obtained by removing one $B$ and one $A$ and adding words which intertwine between the $a$ and $b$ brane indices. This is,  objects that have the index structure of $AB$ and $BA$, but with the ends restricted so that
the corresponding state (operator) does not factorize.

In the end we see that we have all the objects we need to have an approximate Hilbert space of string states with all possible endings, and that these are an approximate Fock space.

A detailed study of these open spin chains and their integrability is beyond the scope of the present paper, but it should be an interesting problem to analyze. 


\section{The chiral ring as a holomorphic quantization of the moduli space of vacua}
\label{sec:cr}

We will now derive the moduli space of vacua of the ABJM theory by analyzing the chiral
ring states classically. Since we are performing classical calculations, the level $k$
of the Chern-Simons Lagrangian will be a real number and not an integer.  Similar considerations
haven been used in \cite{B, Bse} to address classically the problem of ${\cal N}=1$ superconformal field theories in four dimensions, a setup that we will follow closely (see also \cite{IndexSCFT, India} for related ideas).

The basic idea of \cite{Bse} is that the BPS constraint is given by an inequality on the set of states $E\geq R$, where $E$ is the energy of the state, and $R$ is the R-charge. This follows from unitarity of the supersymmetry algebra.
The functions $E,R$ are functions on a phase space. Since $E-R\geq 0$ for all states in the quantum theory, this statement is also true for any classical configuration in the classical limit: it defines a positive operator on the Hilbert space of states. 

The function $E$ as a function on the phase space defined by the field theory is a Hamiltonian function. We can say the same about $R$. The corresponding operators commute in the quantum theory. Thus their corresponding Hamiltonian vector fields on the phase space also 
commute. 
 The set that saturates the BPS  inequality is the set that minimizes the difference $E-R$. In the quantum theory, the chiral ring is a particular set of wave functions localized on the configurations $E-R=0$.

One can easily prove that, for these configurations with $E-R=0$, the two Hamiltonian vector fields associated to the functions $E, \,R$ must coincide (all the details are in \cite{Bse} and the analysis does not change). Thus, for these states it must be the case that the time evolution of the field configuration is given by the R-charge evolution of the corresponding configuration. In the gauge $A_0=0$, this evolution is simply given by 
\begin{equation}
\dot \phi= i R_\phi \phi \label{eq:R}
\end{equation}
for any field with fixed R-charge, $R_\phi$. This equation can be integrated readily.

The only new piece that one has to consider in this case is the Chern-Simons Lagrangian. However, since this  is  first order, the Legendre transform of the action vanishes (because of the antisymmetry) and it does not contribute to the evaluation of the Hamiltonian. 

This permits the computation of the R-charge and the Hamiltonian for such a field configuration, and the comparison of the two field values. Once the equation of motion is used,  the difference between $H$ and $R$ can be written as a sum of squares, all of which must vanish.

We easily  find this way that the configurations must be covariantly constant on the sphere. 
Also, we find that the scalar potential must vanish. The last piece of information that we need to impose is the gauge constraint (the equation of motion of $A_0$). 

Vanishing of the scalar potential implies vanishing of the F-terms. Let us analyze this in 
detail.

The superpotential is given by a term proportional to
\begin{equation}
W= \tr( A_1B_1A_2B_2- A_2B_1A_1B_2)
\end{equation}
Consider a $U(N_1)\times U(N_2)$ gauge group. It is convenient to introduce two auxiliary vector spaces $V_1,\,V_2$ of dimensions $N_1 ,\,N_2$
(acting as the fundamental representation of $U(N_1), \,U(N_2)$ with their corresponding inner product) so that $A, \,B$ act as matrices going from one vector space to the other one. 

Vanishing of the F-terms implies that 
\begin{equation}
(A_a B_{\dot a}) (A_b B_{\dot b}) = (A_b B_{\dot b})(A_a B_{\dot a})
\end{equation}
where we are using matrix multiplication all along.

If we identify the variables $Z_{a\dot a} = A_aB_{\dot a}$, then all of the matrices $Z_{a\dot a}$ commute with each other as matrices going from $V_1$ to itself. Therefore they can be diagonalized simultaneously by a change of basis. Also, one finds that
the matrices $\tilde Z_{\dot a a}= B_{\dot a} A_a$  have similar properties, and moreover they have the same spectrum as the $Z_{a\dot a }$. The proof is as follows:
assume that $Z_{a\dot a} \ket{a} = \lambda \ket a$. Then we have that 
$$ \tilde Z_{\dot a a} (B_{\dot b} \ket a) = B_{\dot b} Z_{a\dot a}\ket a =B_{\dot b} \lambda \ket a= \lambda(B_{\dot b} \ket a)$$
so that $B_{\dot b}\ket a$ is an eigenvector of $\tilde Z_{\dot a a}$ if $\ket a$ is an eigenvector of 
$Z_{a\dot a}$.

One also finds that
the $A,\,B$ matrices intertwine the $Z, \,\tilde Z$ (this structure can be better understood in terms of having an algebra obtained from the superpotential \cite{Brev}, where the matrices $Z_{a\dot a}\oplus \tilde Z_{\dot a a}$ are in the center).  This establishes a non-canonical isomorphism 
between the eigenspaces of the $Z, \,\tilde Z$. This is non-canonical because if we think of the matrices as 
maps between the spaces with their spectral decomposition into eigenvalues, there is freedom to choose phases in the basis representation of the eigenspaces of $Z, \,\tilde Z$. Also, if we have that the rank of the two $U(N)$ factors are different, then the matrices $Z$ and $\tilde Z$ have different rank.

The other square in the potential (in the analysis of \cite{Benna:2008zy} this is called the D-term potential) looks complicated, but it is easy to write is as
\begin{equation}
V_D= \sum_A \tr| \sigma_2 A-A\hat \sigma_2|^2+\sum_B\tr| \sigma_2 B- B\hat \sigma_2|^2\,,
\end{equation}
where the $\sigma$'s are auxiliary fields \cite{Benna:2008zy}.
After all, in the Lagrangian with auxiliary fields we can remove the set of linear terms in $D$ from the potential, because $D$ acts as a lagrange multiplier enforcing a constraint. The fields $\sigma^2, \hat \sigma^2$  would be related to the fourth component of a gauge field from reducing from four-dimensional field theory down to three. 
Setting the potential to zero we get that
\begin{equation}
\sigma_2 A-A\hat \sigma_2 = 0\label{eq:split}
\end{equation}

These equations imply that
$\sigma_2$ commutes with the $Z_{a\dot a}$ and therefore can be simultaneously diagonalized with $Z$ as well . Remember $\sigma$ is hermitian, so we can diagonalize $\sigma$ by  simple $U(N)$ transformations. Thus, the process of diagonalization of the $Z_{a\dot a}$ matrices is done entirely within the set of unitary matrices and this is an allowed gauge transformation, and moreover, the eigenvectors are orthogonal. There is one caveat to the above statement. This is the case as to what happens when $\sigma$ has degenerate eigenspaces. We can guarantee then that $Z$ only has upper triangular form (these configurations should be though of as bound states of branes \cite{BD}). 

If we look at equation (\ref{eq:split}) carefully, we understand that $\sigma\oplus \hat \sigma$ extends 
the center of the corresponding algebra to have one more variable. Moreover, the fields $A,B$ intertwine
the actions of $\sigma, \hat \sigma$.

This means that there is a reduction of the gauge group to $U(1)^N$ on each such vector space if both ranks are the same, otherwise we get a reduction to $U(1)^{N_1}$ and $U(1)^{N_1}\times U(N_2-N_1)$. The intertwiner condition reduces the gauge group further 
to $U(1)^{N_1}\times U(N_2-N_1)$.

Essentially, this tells us that the matrices $A,\, B$ are diagonal (extended by zeroes if the matrices are not square). The diagonal entries of $A,\, B$ parametrize the moduli space of vacua. Thus, the classical solution of the problem tells us that we can make a diagonal 
ansatz and that everything reduces to eigenvalues.

At this stage, we get that the moduli space is covered by $(\BC^4)^{N_1}/S_{N_1}$.
These are the eigenvalues of the matrices. Permutations of eigenvalues are a gauge symmetry, so we get classically the division by $S_{N_1}$. In this $\BC^4$, we still have a $U(1)$ gauge action by phase rotations of the $A,\,B$, moreover we still need to impose the gauge constraint.
Then we get that
\begin{equation}
k F_{\theta\varphi} = Q \label{eq:constraint}\, ,
\end{equation}
and similarly for the $U(N_2)$ fields. 
Here $Q$ is the gauge charge of the field configuration, which is proportional to $|A|^2-|B|^2$
for each eigenvalue and zero on the off-diagonal, for states satisfying the equation of motion
(\ref{eq:R}). Notice that these expressions would be usually associated to solving the D-term
constraints. Also, $Q$ does not have to vanish.

At the classical level if $Q$ does not vanish, we should keep it in the description of the moduli space. The same is true for the canonically conjugate variable, which is the angle of the global gauge transformation.

Since $F_{\theta\varphi}$ is quantized topologically (as we have discussed in Section \ref{sec:dimred}), the set of allowed solutions for each eigenvalue are 
the gauge orbits of the corresponding level sets for equation (\ref{eq:constraint})
is a disconnected non-trivial manifold $\cal M$ with $\pi^0({\cal M})=\BZ$.
 
If we now decide to keep $Q$ and the gauge transformation parameter as variables that describe the moduli space, we can ask what happens when we quantize a diagonal ansatz.
It turns out that the set of diagonal solutions to (\ref{eq:R}) inherits a Poisson structure from  the full dynamical system. In this system
\begin{equation}
\Pi_A = \dot {\bar A} \sim \bar A\,,
\end{equation}
so that the complex variables describing the moduli space 
become canonically conjugate to their complex conjugate variables. Thus, upon quantization, we get a canonical polarization of the Hilbert space where the wave functions are polynomials in the $A, B$ variables.
Moreover, since for the diagonal ansatz the system is free, $A,\,B$ become raising operators
for a Harmonic oscillator, and $\Pi_A\sim \partial_A, \, \Pi_B\sim \partial_B$ represent the lowering operators.

The compatibility between the classical quantization of $F_{\theta\varphi}$, and the quantization of $Q$ in quantum mechanics implies that $k$ is an integer. This is another way to see why the level of the Chern-Simons needs to be quantized. Moreover, this quantization condition restricts the
polynomials in $A,\,B$, so that a monomial wave function  $A^{r_1} B^{r_2}$ is allowed 
only if $r_1-r_2$ is a multiple of $k$. This can be rephrased by saying that we have a holomorphic quantization of $\BC^4/\BZ_k$ for each eigenvalue, as only the $\BZ_k$ invariant wave function on $\BC^4$ are allowed. Indeed, $(r_1-r_2)/k$ becomes the gauge flux on each $U(1)$ of the reduction.
This same reasoning can be extended to other theories in order to compute the moduli space
of vacua, like the orbifolds described in \cite{Benna:2008zy}.\footnote{The moduli space problem for theories with group $U(1)^m$  were also considered recently in \cite{MS}. Our way of doing calculations coincides with their results, but it can also work in other setups.} 

The states with gauge flux should be considered as part of the non-perturbative set of states of the string theory. The extra quantum number is of rotations on the fiber of the Hopf fibration. The simplest objects that carry those quantum numbers in the string limit are D0-branes. Since configurations with flux break the gauge symmetry, it is easy to identify
the degrees of freedom that correspond to ends of strings on the D0-brane: those are the degrees of freedom whose gauge quantum numbers sit in the sector with flux. It should also be interesting to study the corresponding open spin chains associated to these strings and check for integrability. We will have more to about D-brane states later on in the paper.

At this stage the most interesting variation on the theme to study is what happens when the two Chern-Simons levels are different, $(k_1,k_2)$. 
We find using this procedure that the Gauss' law constraint (\ref{eq:quant}) forces us to consider situations where the magnetic flux on the two gauge groups are different. This would imply that the field $A, \, B$ are not scalars on $S^2$, but that instead they have non trivial quantum numbers associated with rotations on the sphere.

 This means that the corresponding state would not be associated to an element of the chiral ring. Also, the quantization condition on the fluxes would tell us that $r_1-r_2$ must be a multiple of both  $k_1$ and $k_2$. A naive guess is that having different Chern-Simons 
terms produces some type of dual theory on $AdS_4\times S^7/\mathbb{Z}_k$, where $k$ is the least common multiple of $k_1, \, k_2$. However, this is misleading.

Another way to address this issue is to realize that $\sigma, \hat \sigma$ are related in a non-trivial way to the fields $A,B$, via the equation of motion of the auxiliary field $D$ in the ${\cal N}=2$ superspace formulation \cite{Benna:2008zy}. They end up being proportional to the charge in these configurations, but with different proportionality constants. Thus the only diagonal solution of \ref{eq:split} is the trivial one, where $\sigma=0$. This also imposes that the gauge flux vanishes, and we obtain instead a moduli space for branes on the conifold geometry: a cone over $T^{1,1}$. The fact that we can have operators that carry an extra D0-brane charge suggests that the eleven-dimensional geometry might be different than a naive $AdS_4\times S^7/\BZ_k$, and instead should be considered as some type of new M-theory fibration over $AdS_4\times T^{1,1}$. This is an interesting topic for future research.

We can also extend our reasoning of giant gravitons to try to understand giant gravitons that include D0-brane charge. It is comforting to know that adding D0-brane charge to a D2-brane giant requires us to 
add flux to the corresponding large eigenvalue. This is required for consistency with the brane within branes understanding of D-brane charge \cite{Douglas}. Understanding how the D0-brane charge gets added to D4-brane giants seems to be somewhat involved and this is beyond the scope of the present paper.

Also, we can notice that in the reduction to eigenvalues, we have to give a multiple of 
$k$ charge to each eigenvalue. Thus, the dybarion determinant operators given by 
(\ref{eq:Dib}) do not exist on their own in the $U(N)$ theory. If the gauge group is $SU(N)$ this might
require a more involved analysis.   However, $\det(A^\dagger)^k$ does exist in the $U(N)$ setup  and has 
mass equal to $k N^2$, suggesting that the operator $\det (A^\dagger)$ carries a fractional charge of $1/k$ that is confined in physical configurations. Also, $\det (A^\dagger B^\dagger)$ does exist, so $\det (B^\dagger)$ carries the opposite charge. Such type of charge seems very reminiscent of situations with discrete torsion and
this suggests that there is some type of discrete torsion for D4-branes.

The operator $\det(A^\dagger)^k$ can also be though of as a bubbling configuration of D0-branes  into a giant, so smaller subdeterminants of $A^k$ with flux should correspond 
to NS5-branes.


\section{Eigenvalue gases}
\label{sec:eg}

As argued in \cite{B, Bse}, to analyze field theories at strong coupling (which necessarily involves $\hbar$), once one has used a dimensional reduction on a sphere, one should also reduce to the moduli space directions. This is stated by saying that the quantum dynamics is dominated by the zeroes of the potential. But if the potential is degenerate, then we end up analyzing the moduli space dynamics instead.

The theories described by \cite{ABJM} have a superpotential that is very similar to the conifold setup \cite{Benna:2008zy}. The minimum of the potential happens when the associated F-terms vanish. As we have seen, these configurations are diagonal.

This gives us a reduction to $U(1)^N$ unbroken symmetry. Also, one would naively imagine that to calculate the moduli space configurations, one would find that it is given by a symmetric product 
$(\BC^4/U(1))^N/S_N$, because various scalar field configurations would be in the same gauge orbit. 
If we used this approach we would find that the moduli space is a real cone over $\CP^3$. $\CP^3$ is the quotient space $S^7/S^1$.
The $U(1)$ gauge orbit acts on the scalars preserving a $SU(4)$ symmetry, and splitting them into 4 complex fields.
This is not the case, as was explained in \cite{DMPV}, but it is useful to keep this picture in mind, because in the string limit one does not get to see the $S^1$ fiber. The difference in treatment is due to the peculiar structure of the Chern-Simons Lagrangian, which is not gauge invariant. Instead of dividing out by $U(1)$, one only divides by $\BZ_k$, where the $\BZ_k$ is a group of large gauge transformations. This was computed
explicitly in \cite{ABJM} and in the previous section.

If we reduce the dynamics classically to the moduli space of vacua, we find a theory of free identical particles on $\BC^4$ with a mass squared equal to $1/4$, and with a left-over dynamics induced from the Chern-Simons Lagrangian.

Upon semiclassical quantization, we find that the angular momentum along the $S^1$ gauge orbits is quantized.  If the  angular momentum were always zero, then the correct description would be indeed that the angular directions form a $\CP^3$. However, the angular momentum along this $S^1$ is allowed to be 
non-zero, as we discussed in equation (\ref{eq:quant}), but it is quantized in multiples of $k$, and not in multiples of $1$. This means that the wave functions at various angles are always the same (in other words one is effectively summing over images). Therefore various points along the $S^7$ get identified. This is how one sees that the moduli space of a single brane goes from $\BC^4$ to $\BC^4/\BZ_k$: the classical quantization condition required by the gauge constraints changes the size of the circle. If we send $k\to \infty$, we recover the naive answer. 

This also explains why the off-diagonal gauge rotations do not add degrees of freedom that one would need to quantize. For these, we do have to impose the gauge constraint, and the off-diagonal charge has to vanish on-shell. 

From the point of view of the quantum mechanics, the wave functions that carry angular momentum on the charge circle of $\BC^4$ are non-perturbative: the gauge field strength is not continuously connected to the trivial configuration and it is quantized at fixed magnetic flux values.

For the special case where we get $\BC^4/\BZ_2$ we have an enhanced $SO(8)$ symmetry of the moduli space wave functions. The new symmetries mix what are naively gauge directions with ordinary R-symmetry directions and they are non-perturbative in nature: they mix the flux quantization with the angular momentum in a non-trivial way.

In summary, so far we have analyzed in detail the classical moduli space and the conditions for naive quantization. What we have found is that the theory can be dimensionally reduced to a one-dimensional matrix model for four normal matrices $\phi^A$, which are commuting with each other and can therefore be diagonalized simultaneously. 

The diagonalization of these matrices can be seen as a gauge rotation which introduces a Vandermonde determinant in the path integral. In order to properly quantize the dynamics of the matrices' degrees of freedom, we need to evaluate this measure, or, in other words, we need to compute the volume of the gauge orbits.
This can be done  by looking at the kinetic term of the matrix model (we follow \cite{Krev,BCott}).

We start by decomposing the matrices $\phi^A$ as follows:
\begin{equation}
\phi^A = U^{-1}\,  D^A \, V \, ,
\end{equation}
where the $D^A$'s are diagonal, and $U$ and $V$ are dynamical gauge transformations that depend on time. These are unitary matrices that can  diagonalize simultaneously all four $\phi^A$'s. 

The kinetic term of the matrix model is given by 
\begin{equation}
\tr\, (\dot \phi^A {\dot { \bar \phi}}_A )\,.
\label{action-kin}
\end{equation}
This can be expanded in off-diagonal infinitesimal angles $\theta^i_j, \, \tilde\theta^{ i'}_{ j'}$, and in the diagonal entries $x ^i_{ i'} \equiv x _i \delta^i_{ i'}$. Taking the matrices $U$ and $V$ to be equal to the identity for $t=0$,  we find that (suppressing for the moment the $SU(4)_R$ index)
\begin{equation}
\dot \phi^i_{ i'} = \dot x ^i_{i'}  - i \dot \theta^i_j x ^j_{ i'}+ i x ^i_{ j'} \dot{ \tilde\theta}^{ j'}_{i'}
\label{decomp-phi}
\end{equation}
It suffices to focus on the off-diagonal angles and to limit ourselves to consider just two eigenvalues, $x _1$ and $ x _2$. We can then write (\ref{decomp-phi}) in the suggestive form
\begin{eqnarray}
\dot \phi = -i \begin{pmatrix}
0& \dot \theta^1_2\\
\dot{  \theta}^{2}_1&0
\end{pmatrix}\begin{pmatrix}
x _1&0\\
0&x _2
\end{pmatrix}
+i\begin{pmatrix}
x _1&0\\
0&x _2
\end{pmatrix}\begin{pmatrix}
0& \dot {\tilde\theta}^1_2\\
\dot{  \tilde \theta}^{2}_1&0
\end{pmatrix}\, .
\end{eqnarray}
The quadratic term in the angle derivatives in (\ref{action-kin}) is then
\begin{equation}
\begin{pmatrix}
\dot{ \theta}^2_1 & \dot {\tilde \theta}^2_1
\end{pmatrix}
\begin{pmatrix}
|x _1|^2 +|x _2|^2
&-2x _1  x^* _2 \\
-2 x^* _1x _2 
&|x _1|^2 +|x _2|^2
\end{pmatrix}
\begin{pmatrix}
\dot \theta^1_2\\
\dot{\tilde \theta}^1_2
\end{pmatrix}\, .
\end{equation}
This kinetic term gives us an induced metric on the angle variables.
The determinant of this quadratic form is the measure factor we are after.
Reinstating the $SU(4)_R$ indices and generalizing to arbitrary $N$, we find in the end that the volume of the gauge orbit $\mu^2$ can be written as a product
\begin{equation}
\mu^2 = \prod_{i<j} \left[( |\vec x _i |^2+|\vec x _j|^2)^2- 4 |\vec x _i \cdot\vec{ x }^{\, *}_j|^2\right]
\label{mu}
\end{equation}
where for each eigenvalue index $i$ we have a four complex vector, $ x ^A_i \equiv \vec x _i$. This measure is gauge invariant, as it should be. 

We should also notice that $\mu^2$ is invariant under rotations of $\vec x _i$ by independent phases. This is a residual gauge transformation that the off-diagonal angles don't see. Some of these rotations by phases of $\exp(2\pi i/k)$ are the identifications induced by the quantization of momenta. 
This result can be contrasted with the calculations that have been done in four-dimensional theories, where one expects a 
measure whose logarithm is a sum over images \cite{BCorr,BCott}. Here we get that the measure is invariant under a larger symmetry group in moduli space.

In case the ranks of the gauge groups are different, there is a similar formula
\begin{equation}
\mu^2 = \prod_{i<j} \left[( |\vec x _i |^2+|\vec x _j|^2)^2- 4 |\vec x _i \cdot\vec{ x }^{\, *}_j|^2\right] |\vec x_i|^{2(N_2-N_1)}\,,
\label{mu2}
\end{equation}
which behaves as if we have $(N_2-N_1)/2$ extra branes at the origin repelling the other branes.
This is a generalization of the Vandermonde measure for rectangular matrices found in \cite{DiF}

The effective Schr\"odinger Hamiltonian for the  moduli space approximation is then given by
\begin{equation}
H_{eff} = \sum_i -\frac 1{2 \mu^2} \nabla_i \cdot \mu^2 \nabla_i +\frac 18 |\vec  x _i|^2\label{eq:heff}
\end{equation}
where the mass term comes from the conformal coupling of the eigenvalues to the curvature of the $S^2$. We emphasize again how this modification of the classical kinetic term by the insertion of the quantum measure $\mu^2$ is due to the change of coordinates from full matrices to eigenvalues, very similarly, for example,  to what happens when one transforms the Laplacian from cartesian to polar coordinates. This result is expanded around a set of classical solutions of the field theory. The measure is a leading semiclassical analysis and should be considered as valid in the free field theory limit only, even though we are assuming that the 'off-diagonal modes are heavy'. Further corrections will come from taking into account how the heavy degrees of freedom are integrated out.

It is easy to see that 
\bea
\psi_0=\exp\left(-\frac{|\vec x _i|^2}{4}\right)
\label{eq:grst}
\eea 
is an eigenfunction of $H_{eff}$, regardless of the precise form of $\mu^2$, so long as $\mu^2$ is a scaling function of the eigenvalues. The precise eigenvalue that one gets depends on the details of $\mu^2$ though.  This is similar to what has been found in previous examples \cite{B, BCorr, BHart}.

It is convenient now to absorb the volume of the gauge orbit into the wave function by defining $\hat \psi_0 \equiv \mu  \psi_0$. With this choice the measure of integration for the hatted wave function is just the standard Euclidean measure $\prod dx_i$. The probability density to find eigenvalues in a given location is given (up to some irrelevant normalization) by 
\bea
|\hat \psi_0|^2 & \sim & \mu^2 \exp\left(-\frac 12 \sum_i |\vec x _i|^2\right)\cr
&=&
 \exp\left( -\frac 12 \sum_i |\vec x _i|^2+ \frac 12\sum_{i,j}\log \left[( |\vec x _i |^2+|\vec x _j|^2)^2- 4 |\vec x _i \cdot\vec{ x }^{\, *}_j|^2\right]\right)\,.
 \label{prob}
\eea
If we go to the thermodynamic limit, we can turn the sums into integrals by introducing an eigenvalue density, $\sum_i \to \int d\vec x \, \rho(\vec x)$. In the four-dimensional version of this story, this has a delta function support as explained in \cite{B,Berenstein:2005jq}. In this case, such an ansatz is given by 
\bea
\rho(x)=N\frac{\delta(|\vec x|-r_0)}{r_0^7 \mbox{Vol}(S^7)}\, .
\eea
The eigenvalues are then distributed uniformly on a sphere of radius $r_0$. This can be guessed by symmetries. The value of the radius can be found by plugging $\rho(x)$ into the expression for the probability and maximizing the exponent. 
One readily finds that 
\bea
|\hat \psi_0|^2 \sim \exp \left( -\frac N 2 r_0^2 +2 N^2\log r_0+ N^2 c\right)\, ,
\eea
where $c$ does not depend on $r_0$. From this it follows that
\begin{equation}
r_0= \sqrt{2N}\, .
\end{equation}

If we take $N_1< N_2$, the two different gauge groups with different rank, we would have 
$N_1$ eigenvalues, but with a modified potential, and the radius becomes $r_0\sim \sqrt{2N_2}$. This is natural also in rectangular matrix models: the Gaussian matrix model  expectation value for $\langle\tr(X X^\dagger)\rangle$ is $ N_1N_2$, which means that each eigenvalue is of order $\sqrt N_2$, as there are only $N_1$ non-zero eigenvalues that one can consider.

This result is exact if supersymmetry cancellations forbid the generation of other terms in the effective potential from integrating out the massive degrees of freedom. These cancellations seem to be present 
if one can argue that all the configurations in moduli space on the sphere  preserve a large fraction of the supersymmetry. 
Also notice that this result has the correct scaling with $N$ near the free field limit, where a Gaussian matrix model would be valid.

In zero-dimensional toy models \cite{BHH}, it has been found that if there are sufficient cancellations 
between bosons and fermions, or if the fermions dominate the calculation, then the scaling like $\sqrt N$ is correct. However, if the cancellations are incomplete, it depends very much on the details. 
What is known in these cases is that the scaling with $N$ is of square root form, but that it can depend on the 't Hooft coupling, and this is a scaling correction. If we call $\lambda $ the combination
$N/k$, the one that is relevant for the planar expansion, we find that we are allowed to set
\begin{equation}
r_0= \gamma \sqrt{N}/\lambda ^\alpha\label{eq:radiusc}
\end{equation}
where $\alpha\geq 0$ and $\gamma$ is a constant numerical factor. 
If $\alpha\neq 0$, this signals some additional strong coupling effects from integrating out degrees of freedom that we have not taken into account so far. 
Since we do not have a complete understanding of the system, it is premature to say how this will work, but we can explore how the dynamics depends on $\alpha$.

Also, as far as counting degrees of freedom of the chiral ring goes, the moduli space approximation captures all of the relevant degrees of freedom.

In particular, as already explained above, if we excite a single eigenvalue by a large amount, we end up in a situation where the system can be analyzed as a single brane motion: all the off-diagonal degrees of freedom connecting
the eigenvalue to the rest of the dynamics become very heavy. This system of a single brane is essentially a free system.

If we give this brane an energy $E$ in a BPS configuration, we find that the radius position in moduli space scales as $\sqrt E$ and that there is no bound in $E$. Such configurations are dual to giant gravitons growing into the $AdS$ direction \cite{GMT,HHI}, see also \cite{BHart} for generalizations. The upshot is that the vevs of the field explore the radial direction in $AdS$. The giant gravitons for the three-dimensional conformal field theories are 2-branes and these are represented by eigenvalues.

A really important question is to determine how the radius scales with $E$ relative to the radius of 
$AdS$. The supergravity calculation tells us that the radius of $AdS$ scales like $N^{1/6}$ in Planck units. Thus, the tension of the brane scales like $(N^{1/6})^3=N^{1/2}$ in $AdS$ units, and thus the typical energy associated to a brane at distance one in $AdS$ units is of order $N^{1/2}$. This corresponds to a radius in field units of 
order $N^{1/4}$. If we compare this typical size with (\ref{eq:radiusc}), we find that the most likely value for $\alpha$ is equal to $1/4$. This suggests that the system receives corrections from strong coupling that are going to be very hard to evaluate. The moduli space approximation needs to be supplemented further with the dynamics of integrating out the off-diagonal degrees of freedom. Notice that if we take the  ansatz for $\alpha=1/4$, then the masses of the off-diagonal degrees of freedom do indeed become very large (they would still scale like $N/k\sim \lambda$) and a truncation to eigenvalues is still justified. However, the effective Hamiltonian should be more complicated than (\ref{eq:heff}).

This renormalization of the radius makes the M-theory geometry much more involved to calculate than the corresponding type IIB geometry in $AdS_5\times S^5$. This should also be apparent in that there is no free fermion picture for M-theory BPS states \cite{LLM}, whereas there is a free fermion picture in the field theory.


\subsection{Masses of off-diagonal degrees of freedom}
\label{masses-sec}

As already remarked in several places, the first thing we should notice is that since the scalar potential of the ABJM theory goes like $\phi^6/k^2$, then the masses squared of the  off-diagonal modes  will necessarily scale like $\phi^4/k^2\sim r_0^4/k^2$ for the vacuum state.
We then expect that the masses of these degrees of freedom will go like
\begin{equation}
m_{od} \sim \lambda^{1-2\alpha} = \frac{N}{k} \lambda^{-2\alpha}
\end{equation}

By contrast, in four-dimensional conformal field theories, one finds that the mass scales differently with $N$: it goes like $\sqrt{g_{YM}^2 N}$.   This is because the $\phi^4$ potential is less steep. If we consider the size $\sqrt N$ as the typical size of the sphere of eigenvalues, a factor of $N$ for the mass indicates that the degrees of freedom
are scaling like an area and not like a length. Also, the fact that the sphere is divided by a $\BZ_k$ symmetry suggests that the object is wrapping the Hopf fibre, since we see a factor of $1/k$ appearing in the effective mass.
This suggests that the off-diagonal modes should be thought of as a membrane wrapping the circle of the Hopf fibre, as already explained in Section \ref{sec:M2vsD2}. 

We proceed now with a careful evaluation of the masses of these off-diagonal modes. The potential for the scalar fields is given, upon reduction on the $S^2$, by (see the appendix for further details on numerical factors)
\bea
V&=&\frac{1}{4}\tr \,\phi^{A}\bar\phi_A-\frac{1}{12 k^2} \left[\tr \, \phi^A \bar\phi_A\phi^B \bar\phi_B\phi^C \bar\phi_C
+\tr \,\bar\phi _A \phi^A \bar\phi_B \phi^B \bar\phi_C \phi^C\right. \cr && \hskip 4cm \left.
+4\,\tr \,\phi^A \bar\phi_B\phi^C \bar\phi_A \phi^B \bar\phi_C
-6\, \tr \,\phi^A \bar\phi_B\phi^B \bar\phi_A\phi^C \bar\phi_C\right]\, , 
\eea
where, as we have seen already, the quadratic term originates from the conformal coupling of the scalars to the curvature of the $S^2$. 

We expand this potential to quadratic order in fluctuations around a background of diagonal matrices. To this scope we write
\bea
(\phi^A)^i_{i'} = x ^A_i\delta^i_{i'}+(\delta \phi^A)^i_{i'}\, .
\eea
After a bit of computations, one finds that the sextic potential contains three different index structures
\bea
V&=&\frac{1}{4}(\delta\phi^A)^i_{j'}(\delta\phi^*_A)^{j'}_{i}\cr && -\frac{1}{4k^2}\Big[
(\delta\phi^A)^i_{j'}(\delta\phi^*_A)^{j'}_{i}\left(
4\, x ^B_i \,x ^*_{Bj}\,x ^C_j \,x ^*_{Ci}-x ^B_i\, x ^*_{Bi}\,x ^C_i \,x ^*_{Ci}-x ^B_j \,x ^*_{Bj}\,x ^C_j\, x ^*_{Cj}
-2\, x ^B_j \,x ^*_{Bj}\,x ^C_i \,x ^*_{Ci}\right)\cr
&&\hskip 1.5cm +
(\delta \phi^A)^i_{j}(\delta \phi^B)^{j}_{i}\left(2 \, x ^*_{Ai}\,x ^*_{Bi}\,x ^C_i\,x ^*_{Cj}
-x ^*_{Ai}\,x ^*_{Bj}\,x ^C_i\,x ^*_{Ci}-x ^*_{Aj}\,x ^*_{Bi}\,x ^C_i\,x ^*_{Ci}\right)\cr
&&\hskip 1.5cm +
(\delta\phi^A)^i_{j'}(\delta\phi^*_B)^{j'}_{i}\left(x ^*_{Aj}\, x ^B_j\, x ^C_i \, x ^*_{Ci}
+x ^*_{Aj}\, x ^B_j\, x ^C_j \, x ^*_{Cj}+x ^*_{Ai}\, x ^B_i\, x ^C_j \, x ^*_{Cj}\right.
\cr && \hskip 4.7cm \left.
+x ^*_{Ai}\, x ^B_i\, x ^C_i \, x ^*_{Ci}
-2\, x ^*_{Aj}\, x ^B_i\, x ^C_j \, x ^*_{Ci}-2\, x ^*_{Ai}\, x ^B_j\, x ^C_i \, x ^*_{Cj}\right)
\Big]\, .
\eea
The first term in the square bracket can be rewritten as
\bea
&&(\delta\phi^A)^i_{j'}(\delta\phi^*_A)^{j'}_{i}\Big(x ^B_i \, x ^*_{Ci}(x ^*_{Bj}\, x ^C_j-x ^*_{Bi}\, x ^C_i)
\cr && \hskip 3cm 
+x ^*_{Bj}\, x ^C_j(x ^B_i\, x ^*_{Ci}-x ^B_j\, x ^*_{Cj}) +2\, x ^*_{Bj}\, x ^*_{Ci}(x ^B_i\, x ^C_j-x ^B_j\, x ^C_i)\Big)\cr
&&\hskip 1cm =-(\delta\phi^A)^i_{j'}(\delta\phi^*_A)^{j'}_{i}\left(\left| x ^B_i \, x ^*_{Ci}-x ^B_j\, x ^*_{Cj}\right|^2+\left| x ^B_i \, x ^C_{j}-x ^B_j\, x ^C_{i}\right|^2\right)\, ,
\eea
whereas the last two terms are given by
\bea
(\delta \phi^A)^i_{j}(\delta \phi^B)^j_{i}\Big(x ^*_{Ai}\, x ^C_i(x ^*_{Bi}\, x ^*_{Cj}-x ^*_{Bj}\, x ^*_{Ci}) +
x ^*_{Bi}\, x ^C_i(x ^*_{Ai}\, x ^*_{Cj}-x ^*_{Aj}\, x ^*_{Ci}) \Big)
\eea
and
\bea
&&(\delta\phi^A)^i_{j'}(\delta\phi^*_B)^{j'}_{i}\Big( 
x ^*_{Aj}\, x ^*_{Ci}(x ^B_j\, x ^C_i-x ^B_i\, x ^C_j)
+x ^*_{Aj}\, x ^C_{j}(x ^B_j\, x ^*_{Cj}-x ^B_i\, x ^*_{Ci})\cr && \hskip 3cm
+x ^*_{Ai}\, x ^*_{Cj}(x ^B_i\, x ^C_j-x ^B_j\, x ^C_i)
+x ^*_{Ai}\, x ^C_{i}(x ^B_i\, x ^*_{Ci}-x ^B_j\, x ^*_{Cj})\Big)\, .
\eea
It is immediate to see then that these last two terms can be eliminated by imposing 
\bea
(\delta \phi^A)^i_{j'}(x ^*_{Ai}\, x ^*_{Cj}-x ^*_{Aj}\, x ^*_{Ci})=0\, , \qquad 
(\delta \phi^A)^i_{j'}(x ^*_{Ai}\, x ^C_{i}-x ^*_{Aj}\, x ^C_{j})=0\, .
\eea
These two conditions amount to a gauge choice and choice of polarization, as explained in \cite{B}.

In the end one obtains that the mass squared,  $m^2_{ij}$, of the off-diagonal mode connecting the $i$-th and $j$-th eigenvalues is given by
\bea
m^2_{ij} &=&  \frac{1}{4}+\frac{1}{4k^2}\left(\left| x ^A_i \, x ^*_{Bi}-x ^A_j\, x ^*_{Bj}\right|^2+\left| x ^A_i \, x ^B_{j}-x ^A_j\, x ^B_{i}\right|^2\right)\cr
&=&  \frac{1}{4}+\frac{1}{4k^2}\left( \left(|\vec x_i|^2+|\vec x_j|^2\right)^2-4|\vec x_i \cdot \vec x^*_j|^2 \right)\, .
\label{mass-od}
\eea

If we consider the theory in flat space, rather than a sphere, we should remove the $1/4$ from the formula, and we should rescale the fields appropriately as described in the appendix (multiplying each $x$ by $\sqrt{4\pi}$).

Notice also how $m_{ij}^2$ has the same dependence on the eigenvalues as the measure of the gauge orbits $\mu^2$ in (\ref{mu}) and is gauge invariant relative to independent phase rotations of the eigenvalues.
This means that the formula can be reduced to a real cone over ${\CP^3}$. We can take the limit
where we consider two nearby branes and ask if this is proportional to the distance between them times the radius. We can then check easily that the mass squared vanishes for the eigenvalues coinciding up to a phase and that it also vanishes quadratically in the difference of coordinates. Moreover it is a scale function. The symmetries guarantee that for small separation of the branes ($\alpha$ and $\beta$ are two constants)
\begin{equation}
\delta m^2 \sim \alpha R^2 (\delta R)^2 +\beta R^4(\delta \theta_{\CP^3})^2\label{eq:metriccp3}\, ,
\end{equation}
so that it is proportional to the distance along $\CP^3$ if the branes are at the same radius. This shows that in the string limit the mass for an open membrane suspended between two membranes behaves like a string suspended between D2-branes. It should also be interesting to study this question of off-diagonal modes in the setup of \cite{Ooguri:2008dk}, where one would expect to see some aspects of the metric of the squashed sphere appearing in the mass formula if the duality conjectures there is correct.

We emphasize again the main difference between the mass of off-diagonal modes in ${\cal N}=4$ super Yang-Mills and (\ref{mass-od}): the former scales like the distance between the two eigenvalues (this is very important  for the analysis of emergent geometry done in \cite{B,BHart}), whereas $m_{ij}$ in (\ref{mass-od}) scales like a distance squared, thus making the off-diagonal modes of the ABJM theory much heavier.


\subsubsection*{Wilson loops revisited}

Having computed carefully the mass of the off-diagonal degrees of freedom, we can be more precise about the definition of Wilson loop operators that we have introduced in Section~\ref{sec:M2vsD2}. Going  back to three-dimensional flat space and reinserting the appropriate factors of $4\pi$ (see the appendix for more details) we find that
\bea
W=\frac{1}{N}\tr \, \mbox{P} \exp\left(\oint iA+\frac{2\pi}{k}\phi^A\bar\phi_B \omega^B_A ds\right)
\eea
where $\omega_A^B$ represents the coupling of the scalars to the loop and can in principle depend on the position on the loop. For the probe brane setup we have used in Section \ref{sec:M2vsD2} this tensor is constant along the loops and has the form
\bea
\omega^B_A=\delta^B_A-2 \hat R^B \hat R_A\, 
\eea 
where $\hat R$ is the unit vector of the distance from the origin of the probe brane.

This is deduced as follows (similarly to what is done in \cite{Malda}). The mass, for a brane at fixed position $R$, and another brane at $\phi$ is given by
\begin{eqnarray}
m^2& =& \frac{4\pi^2}{k^2}\left( \left(|R|^2+|\phi|^2\right)^2-4|\vec R \cdot \vec \phi^*|^2 \right)\cr &
\simeq&\frac{4\pi^2}{k^2}\left(
|R^2|^2 +2 |R|^2|\phi|^2-4|\vec R \cdot \vec \phi^*|^2+\dots\right)\end{eqnarray}
so that when we expand the value of $m$ in powers of $R$, we get
\begin{equation}
m \simeq \frac{2\pi}k\left(|R|^2+|\phi|^2-2 (\hat R\cdot \vec \phi^*)(\hat R^*\cdot \vec \phi)\right)+o(1/R)  
\end{equation}
The term $|R|^2$ is interpreted as the bare mass. The other piece is the leading contribution to the mass from interacting with the field theory degrees of freedom. This is quadratic in the fields and not linear as in ${\cal N}=4$ SYM. The tensor structure of $\omega^B_A$ follows.

It would be interesting to analyze the supersymmetry properties (if there are any) of these objects, looking for example if there are any choices of $\omega_A^B$ which preserve some of the supercharges or cancel UV divergences, as it happens in ${\cal N}=4$ SYM \cite{Drukker:1999zq}. 

This would be related to some type of BPS central charge carried by the fundamental degrees of freedom. In four dimensions one can relate such a possible central charge in a ${\cal N}=2$ system to the electric and magnetic charges of a degree of freedom \cite{Witten:1978mh}. This usually does not work in three dimensions, as electric charges are confined in ordinary Yang-Mills theories. However, a topologically massive Chern-Simons Lagrangian might allow such an interpretation if the asymptotic fields generated by the defect are normalizable at infinity: they have finite energy after a short distance UV cutoff (determined usually by the bare mass) is chosen. 

Another interesting question would be to see if one can compute to all order in perturbation theory
the Wilson loop expectation value for some specific choice of contour, similarly to the circular loop case of ${\cal N}=4$ SYM \cite{ESZ}. This goes beyond the scope of the  present paper and we leave it as an open question.


\section{The strong coupling geometric objects dual to the chiral ring}
\label{sec:strongcoup}

At strong coupling, as we have argued, the chiral ring is described as a holomorphic quantization of the moduli space of vacua. Thus, all interesting wave functions are polynomials in holomorphic variables, times some ground state wave function, that at weak coupling is given by (\ref{eq:grst}). However, a better way to describe the chiral ring is in terms of a measure $\mu$, so that an inner product can be defined between polynomials as follows
\begin{equation}
\langle \alpha|\beta\rangle = \int_{\cal M} d\mu \ \bar\alpha\, \beta \,.
\end{equation}
Here $\mu$ is symmetric under permutation of eigenvalues,  $SU(4)$ invariant, and 
should behave like $\exp(-|\vec x|^2/2)$ on each eigenvalue at large $|\vec x|$. We expect that at strong coupling the saddle point for $\mu$, with $\alpha,\,  \beta$ being the trivial polynomial, will occur for a distribution of eigenvalues that is spherically symmetric and where the eigenvalues are located all at some fixed distance from the origin (this is just like what we calculated at weak coupling). The higher loops should 
correct the measure factor, but not affect the quadratic potential for the effective 
Hamiltonian.

Notice that the saddle describing the distribution of (\ref{prob}) would appear in the action quadratically, with a linear piece (the first term with a single sum) associated to the confining potential and the quadratic piece (the term with a double sum) encoding the logarithmic repulsion of the eigenvalues. These two forces are competing and in equilibrium one has a balance between them.

The linear piece should be non-renormalized. This is the statement that for large field
configurations the off-diagonal modes become heavy and they should decouple from the low energy dynamics, which resembles the classical theory (where one has no Vandermonde repulsion). The wave function should then decay like it does in the harmonic oscillator. 

Now, let us consider various wave functions that can have a simple description. These are the 
wave functions we considered in Section \ref{sec:ffBPS} and were given by traces, totally symmetric Young tableaux, and subdeterminants. It is usually convenient to write them in terms of generating functions. 

We start with the generating function associated to a trace $\tr(A_1 B_1)^J$, which is given by
\begin{equation}
\alpha = \exp\left(\sum_J a_J \tr (A_1B_1)^J\right) \simeq \exp \left(\sum_J a_J \int \rho(x)  (x^1 x^*_3)^J\right)\equiv \exp\left(\int \rho(x) f(x)\right)
\end{equation}
where the $a_J$ are at this stage formal parameters and we have taken $N$ to be large and replaced sums by integrals. Here $x^1$ and $x^*_3$ are the eigenvalues of $A_1$ and $B_1$, respectively. 

We need to find the saddle of $|\alpha|^2 d\mu $ as a function of the distribution
of eigenvalues.
The 
deformation we have studied produces an additional force along the gradient of $f(x)$. If we fix a single spherical harmonic, we can expand to linearized order in variations of $\rho(x)$, and we find that the distribution will be deformed along some spherical harmonic on the sphere.

If we take $J$ large, the deformation is peaked heavily where $x^1 x^*_3$ is maximal. This defines
a particular locus on the distribution where $|x^1|=|x^3|= r_0$. This is a one-parameter family of great circles on $S^7$ that are related to each other by translations along the Hopf fiber. 
When we reduce this family to the $\CP^3$ base, they correspond to a single geodesic of
$\CP^3$. 

We can now consider giving a lot of energy to a single eigenvalue.  These configurations are well described by a single trace, so that we consider instead the generating function given by
\begin{equation}
\alpha= \tr \oint \frac {dz}{2\pi i z}\exp( t  z A_1) \exp(t z^{-1}B_1)\, ,
\end{equation}
where we have included a projector onto the sector where the powers of $A_1$ and $B_1$ are the same.
This is represented by the contour integral in the complex plane $z$ that loops once around zero. 

Given $\alpha$, we can see that for large fixed $t$, and $z$ fixed to be unitary,  a single eigenvalue becomes very different from the rest and can be treated independently. The force on that eigenvalue gets a linear contribution in $t$ and a quadratic force of confinement. The repulsion from the other eigenvalues becomes subleading. These two forces balance each 
other and we find that the single eigenvalue gets large and is removed from the distribution
(this was shown to happen also in \cite{BHart}). Thus we find that the set of states corresponds to an eigenvalue where $|x^1| =|x^3|$, but their relative phase is undetermined (summed over: this is the role of $z$). The upshot is that this represents a single delocalized eigenvalue along the Hopf fiber, but at a fixed position in $\CP^3$, along the same geodesic as before, but at a different radial position. This is the situation that clearly corresponds to a situation where the gauge group is broken classically to $U(1)\times U(N-1)^2$ by going to the Coulomb branch for a single eigenvalue.

Finally, we can consider a generating function for subdeteminants. This is nothing other than a determinant itself. Thus, we can consider the case where
\begin{equation}
\alpha = \det( A_1B_1-s) = \exp\left(\tr\log( A_1B_1-s)\right) \simeq \exp \left(\int \rho(x) \log(x^1 x^*_3-s)\right)\,.
\end{equation}

In the inner product $|\alpha|^2d\mu$ shows that the determinant produces a repulsion of 
all the eigenvalues from the locus $x^1 x^*_3=s$. This is a holomorphic locus that can intersect
the sphere $S^7$ for $s$ less than some value of order $r^2_0$, the radius of the 
distribution. There are giant gravitons that sit on the same locus. Such geometric description of giants has been given by Mikhailov \cite{Mikhailov:2000ya}.

Notice moreover that when we set $s=0$, the determinant factors and the corresponding hypersurface are singular. This corresponds to two D4-branes in the geometry, located at
$x^1=0$ and $x^3=0$ intersecting at right angles (this matches with our counting of string degrees of freedom in Section \ref{sec:ffBPS}).

The parameters $s, \, t,\,  a_J$ give each different notions of coherent states, and they
represent different types of collective coordinates. We see that geometrically they give rise to a Higgs branch ($t$),  a giant graviton ($s$), or to eigenvalue density waves ($a_J$).
One can easily generalize the $s,\, t$ parameter so that they include multiple giants.

Now, let us consider a single D0-brane. At the level of eigenvalues, we are giving flux to one 
eigenvalue in each $U(N)$. These fluxes are the same. At the quantum level this requires
the quantum state expanded around the configuration to have charge $k$. This is, we need to excite $k$ oscillators in the background configuration, that are tied to the eigenvalues with flux.

This operator can be written schematically as $(A^k)_{11}$. The subindex structure indicates the flux amount in the gauge group eigenvalue. If we maximize the angular momentum
relative to the $SU(2)$ R-charge, we find that we should consider $A_1^k$, and that this has 
$SU(2)$ spin $k/2$. If we work in an $SU(4)$ invariant formulation, we have a highest weight state of $SU(4)$ in the $(\ell, 0, 0)$ representation.

The corresponding eigenvalue in the eigenvalue wave function we have discussed so far will have an amplitude that localizes on the sphere, on the region where $|A_1|$ is maximal. This is a point in $\CP^3$, just as is befitting for a D0-brane.
We can also place $\bar{\mbox{D}}$-branes in some location by considering $(\bar A_1)^k_{-1,-1}$, or 
$(B_1)^k_{-1,-1}$.

Given two nearby D0-branes, and let us say that the second is characterized by the replacement $A_1\to \cos\theta A_1+\sin\theta A_2$, we can consider a state with strings stretched between them. These follow from replacing one of the letters $A_1$, and $\cos\theta A_1+\sin\theta A_2$ by a pair of off-diagonal excitations,  say $\bar B_1^\dagger$.

The $B_1^\dagger$ will cost energy proportional to the distance between the localtion of the branes. This follows from the mass formula (\ref{mass-od}) and the comments that lead to (\ref{eq:metriccp3}). Moreover, we can give it some angular momentum on $AdS$ by considering the various spherical harmonics. We get a relation of the form
\begin{equation}
E_{string} \sim \sqrt{ \left(\ell_1+\frac 12\right)^2+ T\delta\theta^2}+\sqrt{ \left(\ell_2+ \frac 12\right)^2+ T\delta\theta^2}-2
\end{equation}
where $T$ will be interpreted as a string tension and it is essentially given by the radius of the eigenvalue distribution. Notice that the two branes carry the same flux, so the off-diagonal modes are expanded in ordinary spherical harmonics.

We can also consider putting a D0/$\bar{\mbox{D}0}$ brane pair near each other.  Then we find that the off-diagonal energy of strings is given by
\begin{equation}
E_{string} \sim \sqrt{  \left(\ell_1+\frac 12\right)^2-1 + T\delta\theta^2}+\sqrt{  \left(\ell_2+\frac 12\right)^2-1+ T\delta\theta^2}-2
\end{equation}
where $\ell_1=1, \dots, \infty$ is the angular momentum carried by the defect. 
This is lower than the energy for strings connecting between two D0-branes. This is evidence that a tachyon should form in the brane/anti-brane system. The tachyon condensation would try to cancel the flux between the branes, and its condensation would move us between different classical saddles, so it should be considered as a non-perturbative effect in field theory. 

Similar processes in the $(2+1)$-dimensional version of M(atrix) theory \cite{BFSS} transferring flux between different eigenvalues 
are also non-perturbative: this is interpreted as momentum transfer in some
T-dual circle in the type IIB string \cite{Berenstein:1997vm}.


\section{Dispersion relation of the giant magnon}
\label{disp-rel-sec}

Equipped with the results from the previous sections, we can now turn to deriving the square root behavior of the dispersion relation for the giant magnon solution found in \cite{Gaiotto:2008cg,Grignani:2008is}. 

This giant magnon is a soliton of type IIA string theory on $AdS_4 \times \mathbb{C}P^3$, which lives in a $\mathbb{R}\times S^2 \times S^2$ subspace, with $\mathbb{R}\subset AdS_4$ and $S^2\times S^2 \subset \mathbb{C}P^3$. This solution rotates with constant velocity along the $S^2\times S^2$ directions, with the same polar angle on the two spheres, $\theta_1=\theta_2\equiv \theta$, but opposite azimuths, $\phi_1=-\phi_2\equiv \phi$.
It is explicitly given in conformal gauge by \cite{Grignani:2008is}
\bea
\theta(\tau,\sigma) =\arccos \left(\frac{\sin \frac{p}{2}}{\cosh u}\right)\, , \qquad  \phi(\tau,\sigma)= \tau +\arctan\left(\tan \frac{p}{2} \tanh u\right)
\eea
where $p$ is the momentum of the magnon, defined by the boundary conditions $\Delta\phi_1=-\Delta\phi_2\equiv p$, and $u\equiv \frac{\sigma - \tau \cos p/2}{\sin p/2}$. Finite size corrections to this configuration were studied in  \cite{Grignani:2008te,Astolfi:2008ji,Shenderovich:2008bs}.

Differently from what happens to its cousin in ${\cal N}=4$ super Yang-Mills \cite{Hofman:2006xt}, the dispersion relation for the $AdS_4 \times \mathbb{C}P^3$ giant magnon turns out to depend on a non-trivial function of the coupling constant $\lambda=N/k$. This fact can be ascribed to the smaller degree of supersymmetry enjoyed by the theory and to the consequent renormalization of the effective action. Combining weak-coupling results \cite{Minahan:2008hf,Gaiotto:2008cg} with the analysis of the Penrose limit of the theory \cite{Gaiotto:2008cg,Grignani:2008is}, and assuming that symmetry arguments along the lines of \cite{Beisert:2005tm} also hold in this setup, the giant magnon dispersion relation has been conjectured to be given by \cite{Gaiotto:2008cg,Grignani:2008is}. 
\bea
\Delta=\sqrt{\frac{1}{4}+h(\lambda) \sin^2\frac{p}{2}}\, , 
\label{disp-rel}
\eea
This is very similar to the result that was computed first in \cite{Santambrogio:2002sb}. Here
 $h(\lambda)$ interpolates between $h(\lambda)=2\lambda$ for $\lambda\gg 1$ and $h(\lambda)=4\lambda^2$ for $\lambda\ll 1$, the perturbative regime, and the dual gravity calculation.

Using the emergent geometry picture that we have developed in the previous sections, we can now prove this square root formula. The emergent geometry approach allows in fact in a very simple way for a complete resummation of certain Feynman diagrams to all orders in $\lambda \sin \frac{p}{2}$, thus giving the exact functional form in (\ref{disp-rel}).

The basic idea is to separate diagonal and off-diagonal modes as, respectively, slow and fast degrees of freedom in a Born-Oppenheimer approximation, and to treat the off-diagonal ones perturbatively \cite{Berenstein:2005jq}. This can be justified on general grounds so long as the off-diagonal modes are sufficiently heavy\cite{BHH}. This means that we turn on the off-diagonal modes $\delta\phi^A$, but we can neglect their back-reaction on the geometry, so that they can be approximated to be free harmonic oscillators with Hamiltonian
\bea
H
=\sum_{i\neq j'} (\Pi_A)^i_{j'}(\Pi^{A*})^{j'}_i + m^2_{ij} (\delta\phi^A)^i_{j'}(\delta\phi^*_A)^{j'}_i \, ,
\label{Hho}
\eea
where $\Pi_A$ is the conjugate momentum of $\delta\phi^A$ and the frequency is given by (\ref{mass-od}).

To compute the energy  of the BMN operators discussed in Section \ref{sec:ffBPS} we consider the corresponding states (recall that the ground state of the spin chain is $\tr(A_1B_1A_1B_1\ldots)$, and $A_1$ has eigenvalues $x^1_i$ while $B_1$ has eigenvalues $x^*_{3 i}$)
\bea
\ket{\psi} \sim \sum_{l =0}^J e^{2\pi iq l/J} \sum_{i,i'}(x^1_i x^*_{3 i})^l (\Phi^\dagger)^i_{i'}( x^{1}_{i'}x^*_{3  i'})^{J-l} (\Phi'^\dagger)^{i'}_i  \hat\psi_0\vac \, ,
\label{psi}
\eea
where we have called $\Phi$ and $\Phi'$ the impurities in the spin chain. As already remarked, they are given by replacing in the combination $(A_1B_1)$ either $A_1$ with $A_2$ or $\bar B_2$ or by replacing $B_1$ with $B_2$ or $\bar A_2$  \cite{Minahan:2008hf,Gaiotto:2008cg}. We  interpret these off-diagonal impurities as raising operators for the Hamiltonian (\ref{Hho}).  We have denoted with  $\vac$ the off-diagonal vacuum. 
The energy of the state $\ket{\psi}$ is given by
\bea
\left< E\right>_\psi=\frac{\int (\prod dx_i)|\hat \psi_0|^2  \sum_{i,i'} \big|\sum_l e^{2\pi iql/J}(x^1_i x^{*}_{3 i})^l (x^{1}_{i'} x^*_{3 i'})^{J-l}\big|^2    \, m_{ii'}}{\int (\prod dx_i)|\hat \psi_0|^2 \sum_{i,i'}\big|\sum_l e^{2\pi iql/J}(x^1_i x^*_{3 i})^l (x^1_{i'} x^*_{3 i'})^{J-l}\big|^2  }\, ,
\label{energy0}
\eea
where we have used that the lowering and raising operators obey 
\bea
[(\Phi)^i_{i'},(\Phi^\dagger)^{j'}_j] = \delta^i_j \delta^{j'}_{i'} \,,
\eea
and similarly for the primed operators.

The integrals in (\ref{energy0}) can be evaluated with a saddle point approximation in two steps. First of all, the integrals are going to be dominated by the configurations which maximize $|\hat \psi_0|^2$, corresponding to eigenvalues that are distributed on the sphere of radius $r_0=\sqrt{2N}$ that we have introduced above. 
We take $N$ to be large, so that the distribution of eigenvalues is continuous.  In this regime, the sum over two particles can be replaced by integrals over the sphere. This is
\begin{equation}
\sum_{i,i'} f(x_i, x_{i'})\to \iint d\Omega_7 d\Omega_7' f(x, x')
\end{equation}
Next, the 
eigenvalue dependence in the sum gives us the amplitude for a particular off-diagonal mode to be excited (remember that eigenvalues are individual positions in the distribution). If we want to maximize this norm, we should be in a locus of the sphere that maximizes $|x^1 x_3^*|$. This is
clearly requires $|x_2|=|x_4|=0$.

We can use the gauge freedom to choose the following convenient parametrization for a pair of eigenvalues that satisfy these constraints
\bea
\vec x=r_0(e^{i\varphi}\cos\vartheta ,\,0,\,\sin\vartheta,\,0)\, ,\qquad 
\vec x'=r_0(e^{i\varphi'}\cos\vartheta' ,\,0,\,\sin\vartheta',\,0)\, .
\eea
Consider now the sum over phases in (\ref{energy0}). The second step of the saddle point approximation consists in maximizing the expressions $(x^1 x^*_3)^l$ and $(x'^1 x'^*_3)^{J-l}$ appearing there. Using the explicit parametrization above, this means requiring that  $\vartheta=\vartheta'=\pi/4$.
We can at this point rewrite the mass of the off-diagonal modes in (\ref{mass-od}) in a more explicit way as
\bea
m^2&=&\frac{1}{4}+\frac{1}{k^2}\left(r_0^4-\left|\vec x \cdot \vec x'^*\right|^2\right)\cr
&=&\frac{1}{4}+4\lambda^2\sin^2\left(\frac{\varphi-\varphi'}{2}\right)\, ,
\eea
where $\lambda=N/k$. 

Putting all together, we get an approximate $\delta$-function for the coordinates that ghave been fixed so far, and we have that the expression for the energy becomes
(for large $N$ the sums over eigenvalues can be replaced with integrals over the relative angles between them) 
\bea
\left< E\right>_\psi=\frac{\int d\varphi\,d\varphi' \, \big|\sum_l \exp\left(2\pi iql/J+i l \varphi +i (J-l) \varphi'\right)\big|^2  \sqrt{\frac 14+4\lambda^2 \sin^2(\varphi-\varphi')/2}}{\int d\varphi \,d\varphi' \, \big|\sum_l \exp\left(2\pi iql/J+i l \varphi +i (J-l) \varphi' \right)\big|^2  }\, .
\label{energy}
\eea
Expanding the modulus squared of the sum over phases one obtains
\bea
\sum_{l,l'=0}^J \exp \left[ i(l-l')\left(\frac{2\pi q}{J}+\varphi-\varphi'\right)\right] &=&\frac{\sin^2(J+1)\left(\frac{2\pi q}{J}+\varphi-\varphi'\right)/2}{\sin^2\left(\frac{2\pi q}{J}+\varphi-\varphi'\right)/2}\cr&\equiv&
\left| \Phi\left(\frac{2\pi q}{J}+\varphi-\varphi'\right)\right|^2\,.
\eea
\begin{figure}[tb]
\begin{center}
\includegraphics[scale=.5]{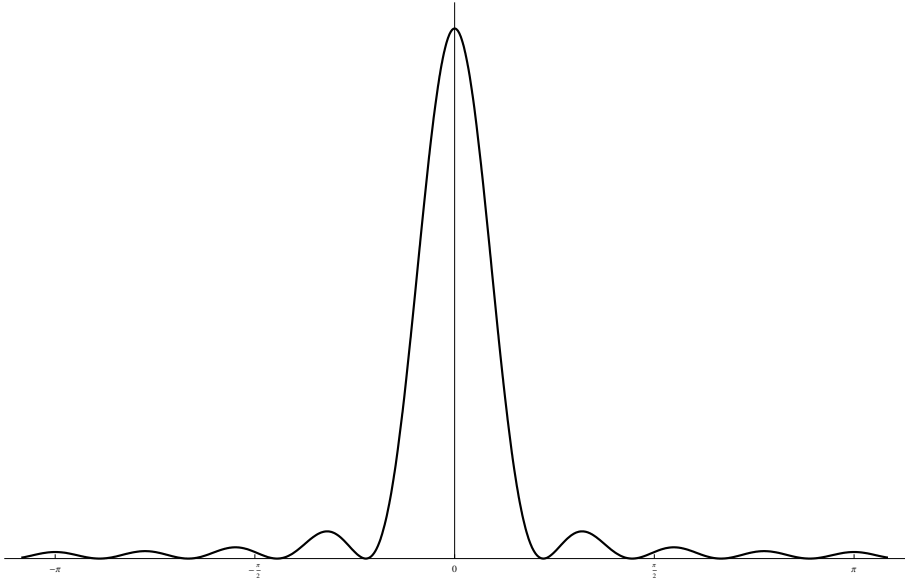} 
\caption{A plot  of the function $\left| \Phi\left(\frac{2\pi q}{J}+\varphi-\varphi'\right)\right|^2$ appearing in the expression for the energy (\ref{energy}). For large $J$ (here we have set $J=8$), this tends to a delta function centered around $\varphi-\varphi'=2\pi q/J$, thus equating the relative phase between two eigenvalues on the sphere to the world-sheet momentum $p$ of the giant magnon.}
\label{phases}
\end{center}
\end{figure}
In the limit of large $J$, this is sharply peaked for $ \varphi-\varphi'=2\pi q/J=p$ (see fig. \ref{phases}), and acts as a delta function. The energy reduces then to
\bea
\left< E \right>_\psi=\sqrt{\frac{1}{4}+4\lambda^2 \sin^2\frac{p}{2}}\, ,
\label{disp-rel-fin}
\eea
in perfect agreement with (\ref{disp-rel}).
We can also consider higher harmonics on the $S^2$, corresponding to bound states of giant magnons \cite{Berenstein:2007zf}, and replace, as already discussed in Section \ref{scalars-sec}, the $1/4$ in the expression above with $w^2_\ell=(\ell +1/2)^2$. This bound state relation is similar to the one found by 
Dorey \cite{Dorey:2006dq}. Giant magnon solutions of the string sigma model also show this result \cite{Ahn:2008hj}.

At strong coupling the radius of the sphere can renormalize and the functional dependence on $\lambda$ inside the square root will change, as expressed in (\ref{disp-rel}).  The emergent geometry picture guarantees nevertheless that the square root behavior will hold to {\it all orders} in perturbation theory in the renormalization of the radius of the distribution and, therefore, also for $\lambda\gg 1$. 

These considerations can be also applied to the fractional M2-brane theory \cite{Aharony:2008gk} with gauge group $U(N_1)\times U(N_2)$. One finds the same  result as (\ref{disp-rel-fin}) except that $\lambda^2$ inside the square root must be replaced by the largest of the coupling constants of the two groups (the radius of the eigenvalue distribution is the square root of the larger $N$). This does not give perfect agreement with the computation of \cite{Bak:2008vd} and suggests that the system is not as simple as in the case where $N_1=N_2$. It should also be noticed that if we look carefully at the hopping of fermion
defects in the spin chain, for $N_1\neq N_2$ we get an inhomogeneous hopping at one loop, on some sites it is proportional to $N_1/k$, and at others it is proportional to $N_2/k$ . This might be non-integrable, but the system might be integrable in the string sigma model limit where no worldsheet fermions are present. Our results are compatible with either possibility.
 A more detailed study is required to settle this issue.


\section{Conclusion}
\label{sec:concl}

In this paper we studied $U(N)\times U(N)$ Chern-Simons theories with ${\cal N}=6 $ supersymmetry and their generalization to groups with different rank, $U(N_1)\times U(N_2)$. We paid special attention to the description of the field theory compactified on $S^2\times \BR$. This geometry is the correct boundary for global $AdS_4$ space and it also plays a fundamental role in the operator/state correspondence for the Euclidean theory.

We studied in detail the chiral ring operators by analyzing states of the field theory on $S^2$. To be elements of the chiral ring we showed that the corresponding classical configurations must be spherically symmetric. We showed in particular that, in the reduction on the sphere, the set of spherically invariant configurations have non-trivial saddles associated to non-zero gauge flux on the sphere. This is quantized already at the classical level and it produces a quantization condition on
the gauge charge that the scalar field configurations carry. We showed that this flux quantization condition only affects the Cartan degrees of freedom. This moduli space is very closely related to the moduli space of vacua of the field theory in flat space. The one difference is the flux quantization condition. If this is removed, then we recover exactly the classical moduli space of vacua of the theory.

At the quantum level, the classical flux quantization provides us with quantized momentum 
along a particular circle fibration of the moduli space of configurations and guarantees that the 
moduli space is an orbifold quotient because the wave functions are restricted. We also were able to compute the moduli space for the case of different ranks for the gauge groups, $U(N_1)\times U(N_2)$, with $N_1<N_2$. This moduli space is the same as that of $N_1$ M2-branes on the quotient space as expected.

We also showed that in the weak 't Hooft coupling limit, where $N/k$ is kept fixed and small, the chiral ring in perturbation theory displays states that can be regarded as the weak coupling description of giant gravitons in $AdS$: these would be given by D2-branes growing on $AdS_4$ and D4-branes growing on $\CP^3$. These states have maximal angular momentum with respect to a $SU(2)\times SU(2)$ sector of the R-symmetry and belong to the $(\ell,0,\ell)$ representation of $SO(6)\simeq SU(4)$. We showed that these states have strings ending on them, in a construction whose combinatorics are very similar to those of ${\cal N}=4$ SYM and we saw that at weak coupling they admit a free fermion description. We also showed that D0-branes could be obtained similarly, but that these configurations are non-perturbative in field theory, since they involve a non-trivial gauge flux. Also, the maximal giant graviton on $\CP^3$ factors into two giants, and we gave evidence for some analog of discrete torsion for these states that forbids a single such brane.

We then studied the system at strong coupling by performing a reduction that works well in four-dimensional systems. We reduced to the moduli space of vacua, which is essentially the set of diagonal matrices, and we tried to solve the corresponding effective dynamics exactly, while treating the off-diagonal modes as a perturbation.

The main calculation was to understand how the volume of the gauge orbit affects the
 eigenvalue dynamics and induces a repulsion between the eigenvalues. The setup is similar to 
 what has been done for ${\cal N}=4 $ SYM before. 
We could show in this way that the eigenvalues form a geometric sphere. However, not all directions along the sphere are the same. The sphere is a natural fibration over $\CP^3$, and the fiber direction behaves very differently than the other directions in the field theory. First of all, it has its origins from a gauge symmetry direction, so the repulsion between the eigenvalues is insensitive to the position of the eigenvalues along this fiber direction. Also, the masses of off-diagonal modes do not care about positions of eigenvalues on the fiber. This means that if we measure distance by the masses of off-diagonal degrees of freedom, then we cannot tell positions apart along the fiber at all. This is not too dissimilar from what happens when M-theory is compactified on a circle and we have the type IIA string description of the dynamics: 
the masses of strings between D-brane states do not know that there are positions of branes along the fiber. After all, the strings are M2-branes wrapped on the fiber,  and they do not have a way to feel that position direction. One needs objects that do not wrap the fiber in order to test the positions of objects along the fiber.

In this sense, the enhancement of the R-symmetry to $SO(8)$ for the case where the Chern-Simons level is $k=1,2$ is extremely
mysterious and requires a much better understanding of the field theory. The arguments that the potential looks $SO(8)$ symmetric are unconvincing, because the gauge dynamics treats the directions very different, and in particular this enhancement is non-perturbative and the configurations that rotate into each other have different amounts of gauge flux. At least in this case, the M-theory geometry is much harder to access than the type IIB geometry dual to ${\cal N}=4$ SYM theory.

A particularly interesting calculation that can be done with this technique is the computation of the giant magnon energies. We were able to prove that the dispersion relation of such a giant magnon (or its bound states) has the form
\begin{equation}
\Delta= \sqrt{ \left(\ell+\frac 12 \right)^2+ h(\lambda) \sin^2(p/2)}
\end{equation}
where $p$ is the world-sheet momentum of the giant magnon excitation, and $\ell$ is the angular momentum on $AdS_4$.  Using a saddle point computation, we showed that $p$ also becomes an angle on the $\CP^3$. The dispersion relation depends on  the function $h(\lambda)$  that we can determine from first principles only at weak coupling. This result is computed independently of any assumption of integrability, and it matches the leading order perturbative calculations from the leading order spin chain model when the ranks of the gauge groups are the same.
 The structure with respect to the quantum number $\ell,\, p$ is exactly as predicted from an all-loop integrability argument and our calculations would show that this structure persists, provided that the approximations that are used in describing the dynamics keep on holding. This provides then very strong evidence for an all orders integrability calculation. 

The quantity $h(\lambda)$ in the field theory is essentially the radius of the eigenvalue distribution determining the ground state. This is not forbidden from receiving perturbative corrections. In contrast, for ${\cal N}=4$ SYM theory, the one loop calculation of $h(\lambda)$ seems to be exact. This has not been explained satisfactorily in the literature:  no one has found  a non-renormalization theorem that would explain this so far. It is possible that this can only occur in the most supersymmetric cases and that it is not applicable anymore in the case of ${
\cal N}=6$ supersymmetry in three dimensions. This renormalization of the radius is required also to match the geometry of giant gravitons with the M-theory dual.

Notice that this same renormalization of the radius that affects the string tension is the one that is responsible for the M-theory specific heat scaling like $N^2/\sqrt\lambda$ (all of them are related to the radius of the $AdS$ geometry in Planck units).\footnote{This same renormalization is responsible for the complicated boundary conditions that one finds in M-theory bubbling geometries \cite{Lin:2005nh}.} It is likely that understanding the radius of the eigenvalue distribution in the vacuum will shed a lot of light on the thermodynamics of the system.

In the end, we are finding that locality on $\CP^3$ can be established easily, especially if we consider that we can relate giant magnon states in the field theory to specific geometric loci, but locality along the Hopf fiber direction is very hard to understand. This seems to require the same type of non-perturbative magic that makes the type IIA string go from ten to eleven dimensions. Remember that this is intuited from the fact that one can have bound states of D0-branes that form a KK tower of gravitons \cite{Witten:1995ex}, but locality on the eleven-dimensional circle is hard to establish. One can postulate it and then check for consistency.  

Considering that the dual CFT is a well defined unitary quantum theory, it might be possible to explore this phenomenon more directly. There is also a possibility of understanding a matrix theory limit \cite{BFSS} of this theory in order to compare with other such constructions.

We have also studied some variations of the ${\cal N}=6$ theory. In particular, for the gauge groups being of different rank, we were able to show that the computation of moduli space produced the theory of $N_1$ branes in the bulk and some branes stuck at the singularity. This matched the interpretation in terms of brane constructions \cite{Aharony:2008gk}, and it provides a first step in the non-perturbative analysis of such setups. We also found that the
giant magnon dispersion relation did not change its functional form in this case either, but we did not find an exact match with the perturbation theory calculation of \cite{Bak:2008vd}. This requires further study.

We also computed aspects of the moduli space for the case where the Chern-Simons levels were different. We showed that some elements of the chiral ring (associated to having non-trivial D0-brane charge) seemed to be missing: the states we found that carried magnetic flux where not in the chiral ring, so we were able to show that the notion of D0-brane charge could still be used. It would be interesting to find out how to think geometrically about these setups.

Overall, we have been able to provide a lot of additional evidence for the AdS/CFT duality conjecture put forward by Aharony {\it et. al} \cite{ABJM} and the integrability of the corresponding string limit. We have shown that the understanding of the geometry of the system is a lot more involved than in the case of ${\cal N}=4 $ SYM. 

We believe that many of the results we found can be extended to other $2+1$ SCFT's, in a similar manner to what has been done in four dimensions \cite{Bse,BHart}. One would have a universal computation of the string spectrum in plane wave limits and of a large part of the supergravity spectrum as well.


\begin{acknowledgments}

D. B. would like to thank S. J. Rey for many discussions related to the material in this paper. We are also happy to thank Sean Hartnoll for discussions.   D. B. is grateful to the Galileo Galilei Institute for Theoretical Physics and to the  Simons Workshop in Mathematical Physics for their hospitality as well as to the INFN for partial support during the completion of this work. This work was supported in part by the U.S. Department of Energy under grant DE-FG02-91ER40618 and NSF grant PHY05-51166.

\end{acknowledgments}


\appendix

\section{Conventions}

In this appendix we collect some of the conventions we have used throughout the paper. 

We use upper case letters from the beginning of the Latin alphabet $A,B,\ldots=1,\ldots,4$ to denote $SU(4)$ $R$-symmetry indices, and lower case undotted (dotted) letters $a,\dot a,\ldots=1,2$ for the first (second) $SU(2)$ factor in the $SU(2)\times SU(2)$ decomposition of $SU(4)_R$, so that the four complex scalar fields of the ABJM theory are $\phi^A=(A_a,\bar B_{\dot a})$. We also define $\phi^A\equiv \vec \phi$. 

We use bars to denote transposed complex conjugates, $\bar {(\phi^A)^i_{i'}}=(\phi^*_A)^{i'}_i$, while we reserve daggers for the raising operators in the Hamiltonian formalism, as, for example, in (\ref{psi}).

Lower case unprimed (primed) indices from the middle of the alphabet $i,i',\ldots=1,\ldots,N$ are the indices transforming in the fundamental representation of the first (second) gauge group factor in $U(N)\times U(N)$. These indices also label the eigenvalues of the scalars.

The action for the scalars that we have used in Section \ref{masses-sec} is obtained from the ABJM action \cite{ABJM}
\bea
{\cal S}&=&\int d^3x\, \tr \Big(D_\mu \bar\phi_A D^\mu \phi^A+\frac{4\pi^2}{3 k^2}\left[\phi^A \bar\phi_A\phi^B \bar\phi_B\phi^C \bar\phi_C
+\bar\phi _A \phi^A \bar\phi_B \phi^B \bar\phi_C \phi^C \right.\cr && \hskip 6cm\left. 
+4\,\phi^A \bar\phi_B\phi^C \bar\phi_A \phi^B \bar\phi_C
-6\,\phi^A \bar\phi_B\phi^B \bar\phi_A\phi^C \bar\phi_C\right] \Big)\cr &&
\eea
via reduction on the $S^2$. In order to have a canonical normalization for the kinetic term, we compensate for the volume of the sphere $\mbox{Vol}(S^2)=4\pi$, by rescaling $\phi\to \phi/\sqrt{4\pi}$.
As discussed in the main text we also need to include the conformal coupling to the curvature of the sphere, whose proportionality factor is given by the square of the dimension of the scalars. We then have 
\bea
{\cal S}&=&\int_{\BR} dt\, \tr \Big(\dot{\bar\phi}_A \dot \phi^A-\frac{1}{4}\bar\phi_A\phi^A+\frac{1}{12 k^2}\left[\phi^A \bar\phi_A\phi^B \bar\phi_B\phi^C \bar\phi_C
+\bar\phi _A \phi^A \bar\phi_B \phi^B \bar\phi_C \phi^C \right.\cr && \hskip 6.5cm\left. 
+4\,\phi^A \bar\phi_B\phi^C \bar\phi_A \phi^B \bar\phi_C
-6\,\phi^A \bar\phi_B\phi^B \bar\phi_A\phi^C \bar\phi_C\right] \Big)\,.\cr &&
\eea
Notice that we denote with the same symbol both the three-dimensional fields and the dimensionally reduced ones.


\end{document}